\documentclass[twocolumn, extraplaces]{aastex7}

\usepackage{xcolor}
\usepackage{graphicx}
\usepackage{appendix}
\usepackage{amsmath,amssymb}
\usepackage{soul}
\usepackage{hyperref}
\usepackage{longtable}
\usepackage[T1]{fontenc}

\received{2026 Sep 18}
\revised{2026 Nov 09}
\accepted{2026 Nov 11}

\begin{document}

\title{Rapid Protoplanetary Disk Evolution and Debris Disk Formation in Taurus--Auriga}

\correspondingauthor{Joshua B. Lovell}; \email{joshualovellastro@gmail.com}

\author[0000-0002-4248-5443]{Joshua B. Lovell}
\affiliation{Center for Astrophysics, Harvard \& Smithsonian, 60 Garden Street, Cambridge, MA 02138-1516, USA}
\email{joshualovellastro@gmail.com}

\author[0000-0003-2253-2270]{Sean M. Andrews}
\affiliation{Center for Astrophysics, Harvard \& Smithsonian, 60 Garden Street, Cambridge, MA 02138-1516, USA}
 \email{sandrews@sao.si.edu}
 
\author[0000-0001-6831-7547]{Grant M. Kennedy}
\affiliation{Malaghan Institute of Medical Research, Kelburn, Wellington 6012, New Zealand}
\email{grantkennedy@gmail.com}

\author[0000-0003-1526-7587]{David J. Wilner}
\affiliation{Center for Astrophysics, Harvard \& Smithsonian, 60 Garden Street, Cambridge, MA 02138-1516, USA}
\email{dwilner@cfa.harvard.edu}

\author[0000-0001-6410-2899]{Richard Alexander}
\affiliation{School of Physics and Astronomy, University of Leicester, University Road, Leicester, LE1 7RH, UK}
\email{richard.alexander@leicester.ac.uk}

\author[0000-0001-9811-568X]{Adam L. Kraus}
\affiliation{University of Texas at Austin, Department of Astronomy, 2515 Speedway, Austin, TX 78712, USA}
\email{alk@astro.as.utexas.edu}

\author[0000-0001-9626-0613]{Daniel M. Krolikowski}
\affiliation{Steward Observatory, University of Arizona, 933 N. Cherry Avenue, Tucson, AZ 85721, USA}
\email{krolikowski@utexas.edu}

\author[0000-0003-4705-3188]{Luca Matr\`a}
\affiliation{School of Physics, Trinity College Dublin, The University of Dublin,
College Green, Dublin 2, Ireland}
\email{lmatra@tcd.ie}

\author[0000-0001-9064-5598]{Mark C. Wyatt}
\affiliation{Institute of Astronomy, University of Cambridge, Madingley Road, Cambridge, CB3 0HA, UK}
\email{wyatt@ast.cam.ac.uk}

\author[0000-0001-5058-695X]{Jonathan P. Williams}
\affiliation{Institute for Astronomy, University of Hawai'i at M$\bar{\texttt{a}}$noa, Honolulu, HI, USA}
\email{jw@hawaii.edu}

\begin{abstract}
We present a Submillimeter Array (SMA) 1.3\,mm (225\,GHz) continuum emission survey of Young Stellar Objects (YSOs) in the Taurus-Auriga star-forming region, comprising 69 class~III and 5 class~I/II YSOs with spectral types B9--M3, as part of a total sample of 149 class~II and 82 class~III YSOs.
This represents a complete sample (100\%) of class~II and III YSOs with spectral types earlier than K3, and a near-complete sample (85\%) for spectral types earlier than M2. The survey has a median $3\sigma$ sensitivity of 1.4\,mJy.
We did not find any emission from circumstellar dust around class~III YSOs, but present three detections that can be attributed to variable stellar activity, and one from an unresolved companion.
We compile and model spectral energy distributions (SEDs) for the sample, and report 27 class~III YSOs with excess thermal emission.
We demonstrate that the class~III YSOs in Taurus are typically older than their class~II counterparts.
We produce millimeter luminosity and dust disk mass distributions, and present evidence for a bifurcation between the class~II and III populations, implying that ${\gtrsim}99.5\%$ of the dust mass is dispersed or transformed as YSOs transition from the class~II to III stage.
The combined infrared and millimeter properties demonstrate that warm (inner) and cold (outer) dust reservoirs evolve on comparable timescales: there is a rapid evolution across the radial extent of disks. 
We demonstrate that the survey properties are broadly consistent with a belt population that formed planetesimals from class~II disks with an efficiency of a few tenths to a few tens of percent.
\end{abstract}



\section{Introduction}
\label{sec:intro}
Young Stellar Objects (YSOs) (young pre-main sequence stars and their circumstellar material) evolve along a sequence that has long been empirically classified from their near- and mid-infrared emission into five classes, referred to as class~0, I, FS (flat spectrum), II and III YSOs \citep{Adams87, Lada87, Greene+1994, Andre+1994}.
These stages are physically attributed to evolution from early to late stage stellar and planetary system formation, as stars accrete dust and gas from their circumstellar material. That material then disperses and/or forms planetesimals and planets, resulting in a significant reduction in their infrared emission \citep[e.g.][]{Williams11}.
Although it is unclear precisely when planetesimals and planets form in this YSO sequence \citep[e.g., see][]{BoothR16,Pinilla16, Carrera17, Nixon18, Carrera20}, there is general consensus that this must occur before the class~III stage, given that class~III YSOs host little-to-no observable circumstellar material which is a pre-requisite for giant planet formation \citep{Drazkowska23}.
Whilst terrestrial planet growth can continue during the class~III stage via collisions between newly formed planetesimals \citep{Morbidelli12, Wyatt15, Su20}, since the primordial dust and gas reservoirs of class~III YSOs has been dispersed or transformed into larger solid bodies, the lack of small solids in these evolved systems substantially diminishes their far-infrared and millimeter emission \citep{Andrews05, Lovell21a, Michel21}.
Disk dispersal in this context combines all processes which diminishes primordial dust and gas reservoirs, and includes gas accretion, gas ejection, dust collisions and radiation pressure blowout, and the subsequent evolution of dust following radial drift.

Complete surveys of class~II YSOs in nearby star-forming regions with the Submillimeter Array (SMA) and the Atacama Large Millimeter/submillimeter Array (ALMA) have contributed significantly to our understanding of disk evolution \citep[see e.g.,][]{Andrews13, Ansdell16, Pascucci16, Ansdell18, Cieza19, Manara+2023}.
Whilst these surveys offer important data on the properties of protoplanetary disks, they suffer from `survivor' bias because they typically exclude class~III YSOs, which are often the dominant fraction of YSOs in the sampled populations \citep[e.g.,][]{Andrews20}.
Analyses of disk properties that exclude the class~III population overestimate the millimeter luminosity and disk mass distributions, and thus bias interpretations of how disks evolve.
This is particularly important when comparing disk properties to more evolved systems that host debris disks and exoplanets.
Whilst steps have been taken to examine the millimeter properties of class~III YSOs and subsequent planetesimal disk evolution \citep{Kenyon+2008, Hardy15, Cieza19, Lovell21a, Lovell21b, Michel21, Najita+2022}, there still remains no complete millimeter survey of class~III YSOs in any star-forming region at sufficient depth.

In this work, we confront this survivor bias and observe a complete sample of YSOs for stellar hosts with a broad spectral type (SpT) range. 
We obtained new SMA data for YSOs in the Taurus-Auriga star-forming complex (herein Taurus), and collated available SMA and ALMA data from the literature, to analyze in the context of protoplanetary disk evolution and debris disk development.
In \S\ref{sec:sample}, we present the sample of Taurus YSOs.
In \S\ref{sec:obs}, we describe the new SMA observations, data reduction, calibration, and flux extraction, the collation of archival SMA/ALMA data, and the generation of millimeter luminosity distributions.
In \S\ref{sec:results}, we present an overview of the key findings from both the millimeter and multi-wavelength analyses, and discuss these results in the context of the physical evolution of YSOs.
We summarise and conclude our work in \S\ref{sec:conclusions}.
In \S\ref{sec:data}, we provide links to SMA continuum images and measurement sets, as well as reduction scripts \citep{Lovell_26_Taurus_data}.

\section{A Sample of Taurus YSOs} \label{sec:sample}
\subsection{Sample selection and properties}  \label{sec:sampleProps}
The YSO targets we observed with the SMA were selected based on their identification in the Taurus-Auriga complex with the Gaia EDR3 catalog presented by \citet{Krolikowski21}, herein K+21.
This catalog contains 587 sources, of which 75 are listed as `non-members' with no clear correspondence to known sub-regions in the complex and so are excluded \citep[this same sample was discussed by][]{Rahman+2025}.
We refer to the remaining 512 sources as the `full K+21 sample'.
In \S\ref{disc:sampleSelection} we considered the impact on our conclusions if instead the more recent Taurus census of \citet{Luhman23a} had been adopted instead.

Of the 512 sources, 507 (99\%) have SpTs collated from the literature by K+21, and comprise four B stars, five A stars, three F stars, fourteen G stars, 83 K stars, and 398 M stars, which range from B9--M9.5.
Five sources (COU-706-B, J2-2041B, HD~27659~B, IW~Tau~B and GZ~Aur) lack SpTs in the K+21 catalog. 
Although COU~706~B and HD~27659~B, lack spectral types in the literature, these were observed by the SMA based on their photospheric properties (see \S\ref{sec:obs}). 
We used Gaia-2MASS colors of the primaries to estimate the interstellar extinction to each of these two systems \citep[based on the color excess above the zero-extinction color-SpT relations of][]{pecault_mamjek_2013}, and then used the same relation to estimate the companion spectral type from its de-reddened color. 
From its Bp-Rp color, we estimate a spectral type of G4 for COU~706~B.
From its G-K color, we estimate a spectral type of G7 for HD~27659~B.
We include these SpTs in Table~\ref{tab:mainTable}.
This resulted in a total sample of 509/512 YSOs with SpTs.
In total, 389/512 sources do not have flags associated with known stellar multiplicity (76\%) from K+21.
We did not attempt a complete verification of the multiplicity flags, but updated five sources by adding: a spectroscopic binary flag (`S') to COU-706-B, a hierarchical triple flag (`Tri') to HP~Tau~G2 and HP~Tau~G3 \citep[part of a triple system, see e.g.][]{Reipurth+2026}, unresolved flags (`U') to both HD~31305 and 2MASS~J04161407+2758275 (which have been interpreted as close binaries in the literature), and resolved flags (`R') to HD~28867~A and B.

\begin{figure}
    \centering
    \includegraphics[clip, trim={0.5cm 0.8cm 2.0cm 2.4cm}, width=0.99\linewidth]{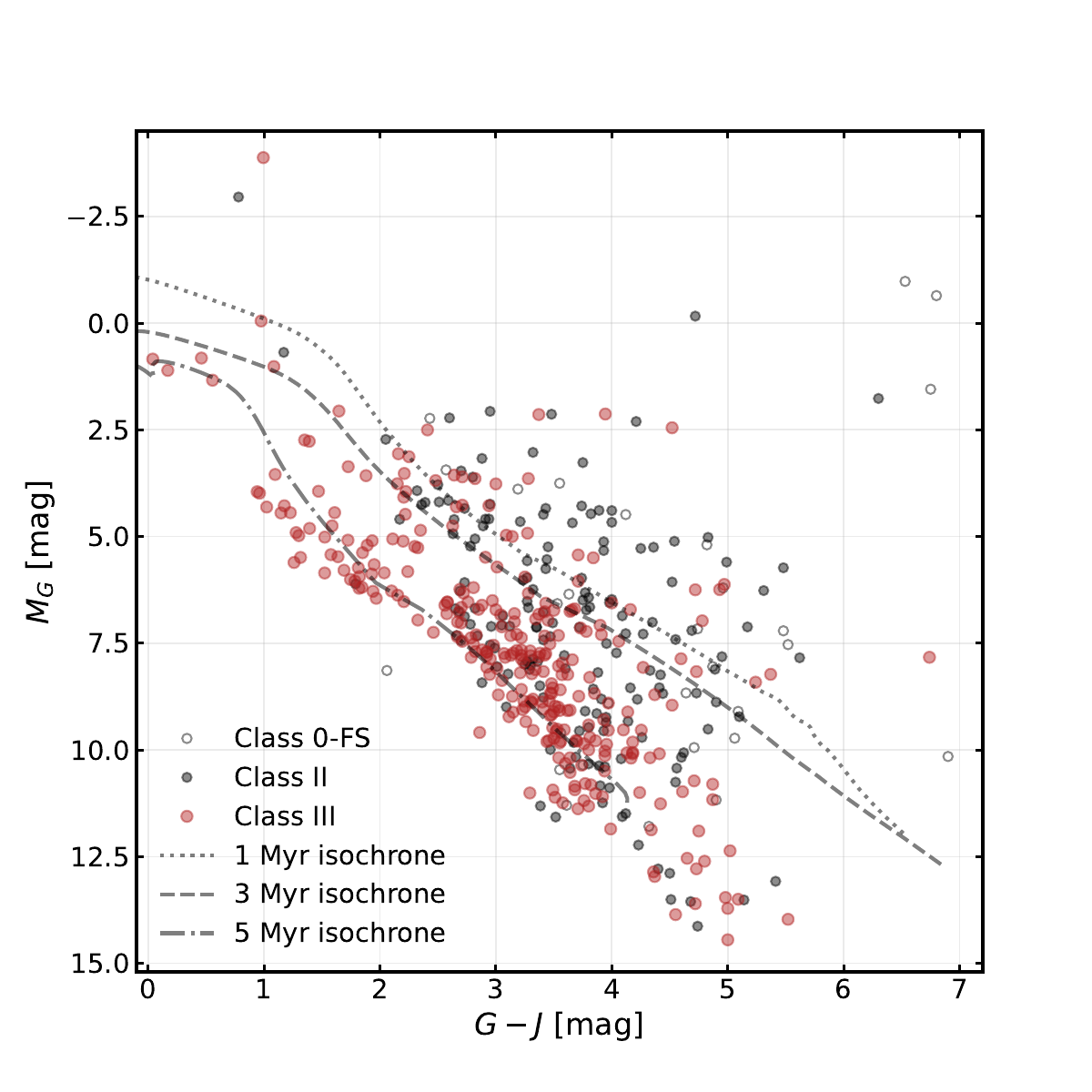}
    \caption{Color-magnitude diagram for the full sample with available Gaia $G$-band and 2MASS $J$-Band magnitudes (484/512 sources total). Markers indicate class 0--FS YSOs (gray circles), class~II YSOs (filled black dots) and class~III YSOs (filled red dots). {\tt PARSEC} isochrones for 1, 3, and 5\,Myr are overplotted.}
    \label{fig:CMD}
\end{figure}

Figure \ref{fig:CMD} presents a color-magnitude diagram of the 484/512 sources in the sample with Gaia $G$-band and 2MASS $J$-band magnitudes, colored by YSO class.
We adopted the extinction-corrected, absolute magnitudes, $M_G$, based on known parallax distances ($d$), and available $G$, $V$, and $J$-band extinction values, i.e., $M_G = G \,+\, 5.0 -5\log{d} -A_G$, utilising simplified extinction approximations of $A_G=0.86A_V$ and $A_G=3.0A_J$ for sources where $A_G$ was not available \citep[see e.g.][]{Cardelli+1989, Danielski+2018}.
The 1, 3 and 5\,Myr {\tt PARSEC} (v1.2) isochrones are overplotted, obtained from the \href{https://stev.oapd.inaf.it/cgi-bin/cmd}{CMD~3.9 web interface} \citep[for details on {\tt PARSEC} v1.2, see][]{Tang+2014, Chen+2015} for Solar-metallicity stars without circumstellar emission, corrected for unresolved binaries, in the mass range $0.1\,M_\odot{<}M_\star{<}16\,M_\odot$.
Whilst these isochrones are only intended to be used for a qualitative comparison, they show broad consistency with the range of derived ages for the Taurus sub-groups in K+21.
There is a visually apparent correspondence between YSO class and age (discussed further in \S\ref{sec:stars}), whereby class~III YSOs typically appear older (closer to the 5\,Myr isochrone) in comparison to the more scattered class~II population that better overlaps with the 1--3\,Myr isochrones.

\subsection{Spectral Energy Distributions and YSO classes} \label{sec:YSOclasses}
The YSO classes of pre-main sequence stars are measured from their infrared spectral slope,
\begin{equation}
    \alpha_{\rm IR} = -\frac{\partial\log{\nu F_\nu}}{\partial\log{\nu}},
\end{equation}
following the classical definitions of \citet{Lada87}, \citet{Adams87}, \citet{Greene+1994}, and \citet{Andre+1994}, where $\nu$ is frequency, and $F_{\nu}$ is flux at $\nu$, typically measured from the near- to mid-infrared.
The precise empirical boundaries between YSO classes have shifted over time: we adopt the definitions from \citet{Williams11}, where class~III YSOs have $\alpha_{\rm IR}{<}{-1.6}$, class~II YSOs have $-1.6{<}\alpha_{\rm IR}{<}-0.3$, class~FS YSOs have $-0.3{<}\alpha_{\rm IR}{<}0.3$, and class~I YSOs have $\alpha_{\rm IR}{>}0.3$.
Since class~0 YSOs are defined as having no optical or near-IR emission, these sources only possess lower-limits on their spectral slopes, and by the above definition appear in the same bin as class~I YSOs.

We constructed spectral energy distributions (SEDs) to consistently measure the spectral slopes, and derive stellar and disk properties for the full K+21 sample.
SEDs were generated by {\tt sdb}, which fits {\tt PHOENIX} stellar atmospheres \citep{Husser+2013} with and without extinction to model stellar emission ($L_\star$, $T_{\rm eff}$) and interstellar reddening \citep[$A_V$, based on][via the python {\tt extinction} package with $R_V=3.1$]{Odonnell+1994}, and where warranted, one or two-component modified blackbody spectra representing single-temperature disk components from which blackbody radii ($r_{\rm bb}$), fractional luminosities ($f$), turnover wavelengths ($\lambda_0$) and spectral slope enhancements ($\beta$) are determined \citep[for a detailed description of the fitting methodology, see][]{Yelverton19}.\footnote{{\tt sdb} is the public window into an SED fitting database developed by Dr.~Grant Kennedy, available at \href{https://sdb.drgmk.com}{https://sdb.drgmk.com}. {\tt sdb} i) collates publicly available fluxes of sources via a script (\href{https://github.com/drgmk/sdb}{https://github.com/drgmk/sdb}) and ii) fits SED models to data (\href{https://github.com/drgmk/sdf}{https://github.com/drgmk/sdf}).}

\begin{figure}
    \centering
    \includegraphics[clip, trim={0.0cm 0.0cm 0.0cm 0.0cm}, width=1.0\linewidth]{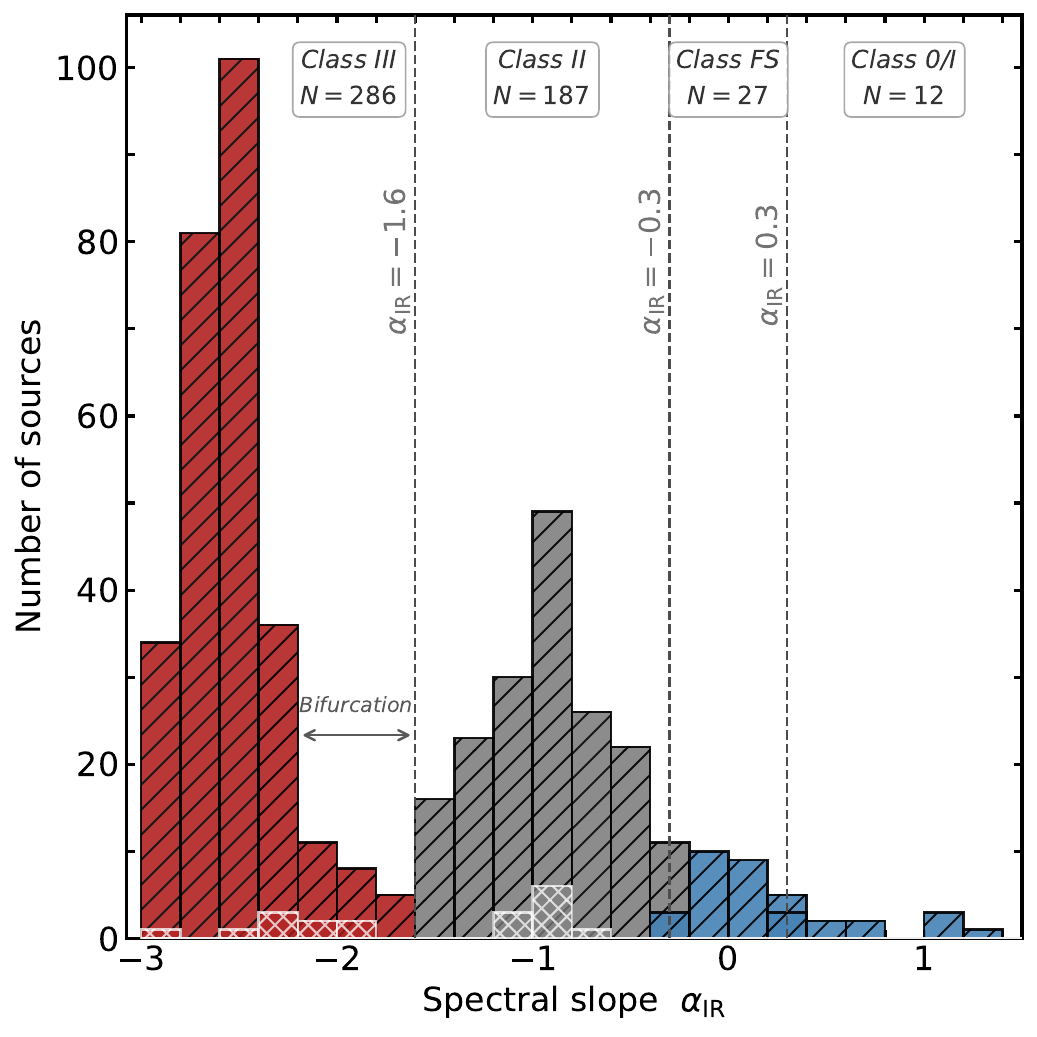}
    \caption{Histogram of $\alpha_{\rm IR}$, where bins show derived values, where colors reflect the YSO class (red for class~III, gray for class~II, and blue for class~0--FS), with dashed-hatching for derived $\alpha_{\rm IR}$ values, and cross-hatching for upper limits. The YSO class boundaries are marked, along with the bifurcation where a significant drop is seen between class~II and III YSOs.}
    \label{fig:alpha_hist}
\end{figure}

We extracted all 512 SEDs using {\tt sdb}.
In general {\tt sdb} collated {\it Gaia}, 2MASS (Two Micron All Sky Survey), {\it WISE} (Wide-Field Infrared Survey Explorer), {\it AKARI}, and {\it IRAS} (Infrared Astronomical Satellite) data for all sources, as well as {\it Spitzer} (IRAC; Infrared Array Camera, and MIPS; Multiband Imaging Photometer) and {\it Herschel} (PACS; Photodetector Array Camera and Spectrometer, and SPIRE; Spectral and Photometric Imaging Receiver) data for a subset of sources.
We imported long-wavelength data from SMA and/or ALMA into {\tt sdb} (see \S\ref{sec:SED_data}) to better model the long-wavelength emission (note that these data had no impact on the assessment of $\alpha_{\rm IR}$).

We assessed all SED fits by-eye and found that sources with little-to-no infrared excesses were systematically well-fitted by {\tt sdb} (i.e. all the class~IIIs), but that earlier stage YSOs with bright infrared excesses and/or more significant optical/infrared obscuration were systematically not well-fitted owing to their intrinsic disk properties (e.g. a broad excess is not always well approximated by two blackbody functions).
For this reason, whilst {\tt sdb} provides reliable data for all sources, and derived stellar (and disk) parameters for class~III YSOs, we do not use the {\tt sdb}-derived stellar and disk model properties of class~0--II YSOs.
We found some sources with incomplete data, which were most often YSOs with very late spectral types and those with `low-quality' data flags, which in some instances biased their model fits.
These are carefully accounted for in the analyses that depend on SpT.
In addition, one source of importance to the millimeter survey (see \S\ref{sec:data}), V1195~Tau, was not resolved by most of the available photometry, resulting in a degeneracy between stellar and disk parameters for the binary stellar components. 
HD~28867B had a similar problem.
For V1195~Tau and HD~28867, we utilise the best-fit stellar parameters of the A-components and map those to the B-components; since HD~28867 is a B9--B9 binary \citep{Walter+2003}, and V1195~Tau is a K6.5--K6.5 binary, we deem this to be a reasonable first-order assumption.

Since {\tt sdb} reliably imported data for all 512 sources, we were able to measure the distribution of $\alpha_{\rm IR}$ directly from the output photometry table.
Figure~\ref{fig:alpha_hist} presents a histogram of $\alpha_{\rm IR}$ values (in red for class~IIIs, gray for class~IIs, and blue for class~0--FS) and upper limits (cross-hatched in red and gray respectively) for the full sample.
This plot presents $\alpha_{\rm IR}$ based on either the measured 2.2--23.7$\mu$m (2MASS $K_s$-band to {\it Spitzer} MIPS-24) data (130/512 sources), the 2.2--22.1$\mu$m (2MASS $K_s$-band to {\it WISE} $W4$) data (166/512 sources), the 2.2--11.6$\mu$m (2MASS $K_s$-band to {\it WISE} $W3$) data (134/512 sources), or where MIPS24, {\it WISE} $W4$ and $W3$ were absent, the model-derived 2.2--23.7$\mu$m slope (82/512 sources).

This histogram demonstrates that $\sim$56\%\ (286/512) of the K+21 sample are class~IIIs, 37\%\ (187/512) are class~IIs, and 8\%\ (39/512) are earlier-stage YSOs (class~0--FS).
Nineteen sources have upper limits on $\alpha_{\rm IR}$, of which ten are counted in the class~II fraction; as such the class~II fraction could be as low as 35\%, and the class~III fraction as high as 58\%.
The histogram also shows a clear bifurcation between the class~II and III populations, in that there are only a small number of class~IIIs with measurable infrared excesses, as expected given the typical distributions of near-to-mid-infrared color-color diagrams of young star forming regions \citep[see e.g.,][for studies of Taurus in particular]{Luhman2010, Luhman23a}.
We reproduced this distribution using only MIPS-24 data and model derived values, and separately using only {\tt WISE W3} data and model derived values, and found the bifurcation persisted in these distinct samples.
To verify this distribution, we compared the derived $\alpha_{\rm IR}$ values with those presented in \citet{Luhman2010}. 
Although there are only 282 YSOs in both the K+21 sample and the \citet{Luhman2010} {\it Spitzer}-based catalog, we found that 276/282 YSOs agreed in the assessments of YSO classes.
The small number of disagreements reflect the slightly different wavelength ranges or datasets adopted.

\begin{figure}
    \centering
    \includegraphics[clip, trim={0.0cm 0.0cm 0.0cm 0.0cm}, width=1.0\linewidth]{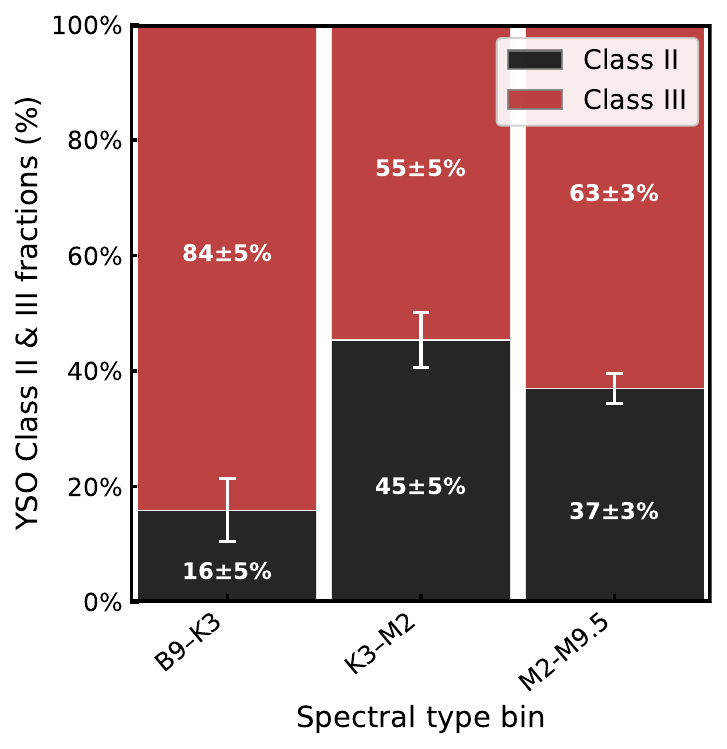}
    \caption{Fractional levels of class~II and III YSOs based on {\tt sdb} models, split into three SpT bins. The error bars mark the 68\%\ confidence levels based on the Wald interval. }
    \label{fig:chart_cii_ciii}
\end{figure}

Although the precise $\alpha_{\rm IR}$ values for class~0--II YSOs derived by {\tt sdb} models are biased, {\tt sdb} models can reliably distinguish YSOs as being a class~III or not.
Figure~\ref{fig:chart_cii_ciii} shows the fractions of class~II and class~III YSOs based on the {\tt sdb}-derived $\alpha_{\rm IR}$ for three SpT bins: early- (B9--K3), mid- (K3--M2), and late-type (M2--M9.5).
For binomial sample uncertainties (estimated from the Wald method with 68\% confidence intervals), the fractional population of class~III YSOs peaks towards the earliest-type (earlier than K3) stars, with a $84$:$16$ class~III-to-class~II ratio, falling to $55$:$45$ and $63$:$37$ towards stars later than SpT K3 (in the K3--M2 and M2--M9.5+ bins respectively).
The fractional enhancement in class~III YSOs is significant, and suggests that disk evolution in Taurus is more advanced for the earliest type stars, corroborating other analyses \citep[see e.g.][]{Kennedy+2009, Pfalzner+2026}.
We verified that these results remain unchanged whether we include/exclude the small number of class~0, I and FS sources with $\alpha_{\rm IR}{>}-0.3$.
Although not included on this figure, we found that finer sample binning revealed a relatively flat class~III-to-II ratio from mid-K to late-M stars, for example the ratios in the K3--K9, M0--M2, M2--M5 and M5+ bins were $65:35$ (57 sources), $43:57$ (51 sources), $62:38$ (161 sources) and $64:36$ (164 sources) respectively.

\section{Observations and Data Reduction} \label{sec:obs}
\subsection{Millimeter observations}
\label{sec:obsSMA}
\subsubsection{SMA observational setup}
The Submillimeter Array (SMA) is an 8-antenna (sub)millimeter interferometer based on Maunakea, Hawai'i \citep{Ho2004}.
We present observations of YSOs collected during 33 observing tracks, in SMA projects 2021B-S014, 2023B-S038 and 2025B-S040 (PI: J. B. Lovell), 2021A-S070 and 2021B-S012 (PI: S. Andrews), and 2021B-H003 (PI: J. P. Williams)\footnote{The SMA observation IDs were as follows: 10871, 10872, 10910, 10943, 10950, 10962, 10991, 11000, 11002, 11003, 11006, 11007, 11010, 11012, 11019, 11021, 11025, 11028, 11045, 11051, 11076, 11078, 11105, 11128, 11135, 11137, 11142, 11145, 11151, 11159, 11178, 12491, and 14137.}.
The observed target fields included 69 class~III YSOs, and 5 class~I--II YSOs (two intentionally, and 3 objects that were in the SMA primary beam half-power-beam-widths of targeted class~IIIs). 
By design, these observing programs preferentially sampled YSOs with earlier spectral types.
All observations had phase centers set to target coordinates (J2000) as in Table~\ref{tab:mainTable}, based on {\it Gaia} EDR3 \citep{Gaia21}.
The observations were conducted in the SMA `Compact’ (COM) configuration, with projected baselines that spanned 5.1--77.0\,m.
The total integrated times-on-source ranged from 32~mins (single `snapshot' observations) and 264~mins (multi-night snapshot observations).
The mean precipitable water vapour (PWV) levels of all observations were PWV${<}4\,$mm.
Observations were conducted with between 5--8 available antennas.

SMA observations were typically scheduled to observe a bandpass calibrator, flux calibrator, and then cycle the `target loop', which includes nearby gain phase calibrators and all science targets.
Many tracks include additional mid-track flux calibration, to account for atmospheric transmission differences at the start and end of observations, and final calibration loops to obtain further bandpass calibration data.
All calibrators were selected from the standard SMA list.
Some tracks were obtained as `filler' time, and thus shared bandpass and flux calibration data with different science projects.
All observations were conducted after the SMA `Wideband Astronomical ROACH2 Machine' (SWARM) correlator \citep[see][]{Primiani+2016J}. 
The mean local oscillator (LO) frequency was 225.5\,GHz/$1.33\,$mm.
For a small subset of filler tracks, the precise tuning was not selected by the PI: these LOs range from 229.5--247.0\,GHz.\footnote{Single-target filler tracks had their LOs tuned to 229.5\,GHz in observations with IDs 10871, 10872, 11000, 11010, 11012, 11019, and 11021, to 231.0\,GHz in 10943, to 239.5\,GHz in 10991, to 239.8\,GHz in 10910, and 247.0\,GHz in 10950.}
Standard flux, bandpass, and gain calibrators were used in all observations, the latter referenced to the target location, and accessed via the SMA public data archive \href{https://lweb.cfa.harvard.edu/sma-archive/}{https://lweb.cfa.harvard.edu/sma-archive/}.

To obtain \textit{Common Astronomy Software Applications} ({\tt CASA}) measurement sets \citep{CASA}, we converted each raw SMA dataset with {\tt pyuvdata} version 3.1.3 \citep{pyuvdata, pyuvdata_Karto2025} with a spectral re-binning factor of 256 (35.84\,MHz), optimised for continuum-only analysis.\footnote{Whilst the primary focus of this survey was to understand the evolution of continuum emission, the setup of the SMA was designed such that for most tracks the $J$=2--1 transitions of $^{12}$CO, $^{13}$CO and C$^{18}$O were included within the spectral bandpasses at a native spectral channel resolution of 140\,kHz (${\approx}0.2\,$km\,s$^{-1}$).}
Auto-flagging, gain (phase and amplitude), bandpass, and flux solution tables were calculated with the SMA {\tt COMPASS} (Calibrator Observations for Measuring the Performance of Array Sensitivity and Stability) tool version 0.11.0 (Keating et al., in prep.), and applied to the science measurement sets via a standard SMA reduction script\footnote{A general version of the SMA reduction script can be accessed via: \url{https://github.com/Smithsonian/sma-data-reduction}. We provide the calibration scripts for each dataset in the companion data repository associated with this publication, see \S\ref{sec:data}.} in {\tt CASA} v6.6.3, utilising the {\tt applycal} task.
Prior to the application of the calibration solutions to the science data, calibrator data was checked per-spectral window and per-antenna for spurious data as a function of channel (frequency) and time. 
Manual flagging was applied to avoid any data that skews or contaminates the science measurement sets.
This routine follows \citet{Lovell+2025a, Lovell+2025b}, where the same calibration procedure was conducted for other analogous analyses.

\subsubsection{SMA images and millimeter fluxes} \label{sec:images_fluxes}
\begin{figure*}
    \centering
    \includegraphics[width=0.9\linewidth]{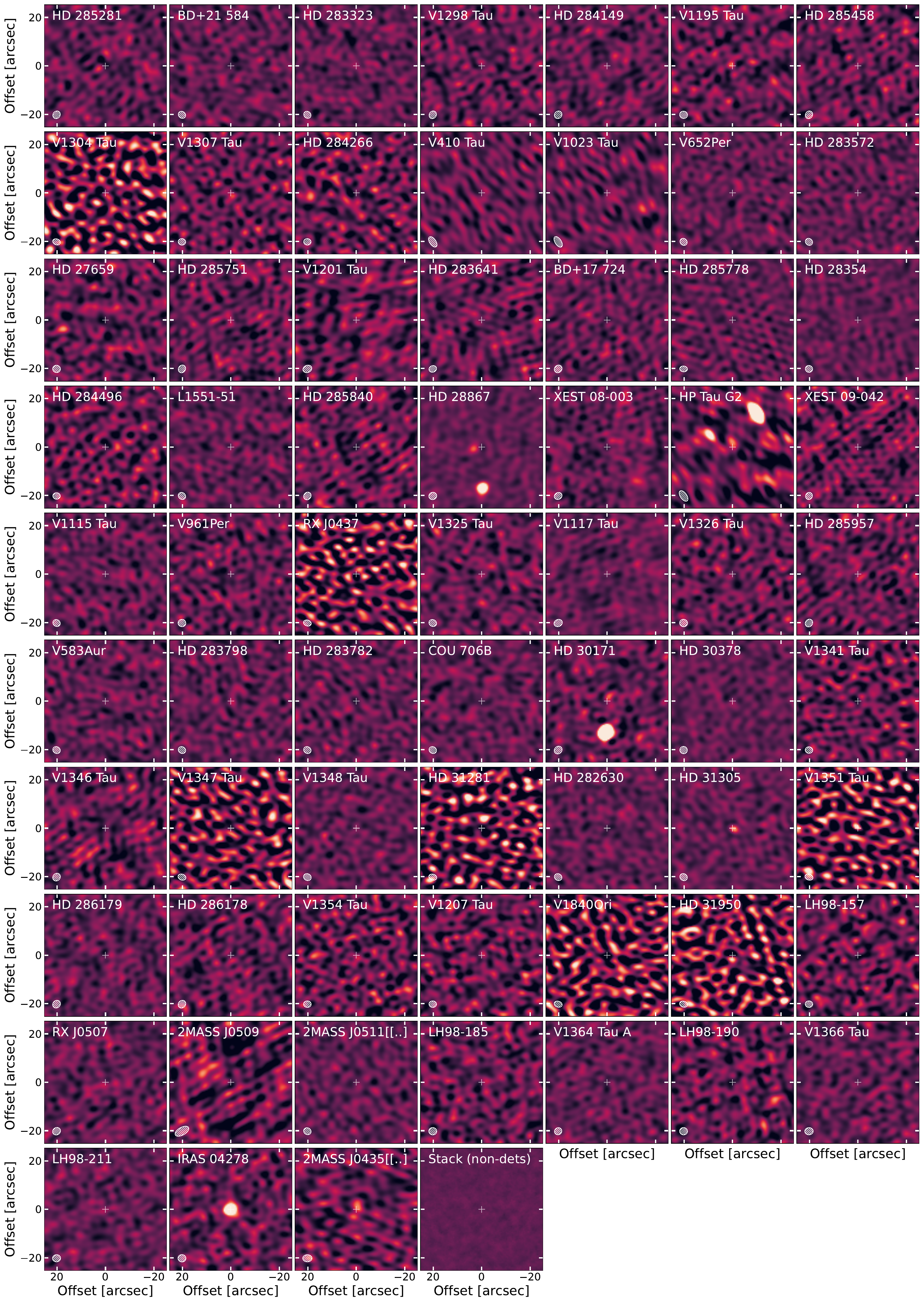}
    \caption{Gallery of SMA continuum images, including in the lower-right of each panel the restored (clean) beams, and white cross-hairs marking the phase center. All images are linearly scaled from -1 to 3\,mJy\,beam$^{-1}$.}
    \label{fig:contgallery}
\end{figure*}

We produced {\tt tclean} images for each target observation, and additionally imaged all measurement sets together for sources with multiple observations.
We ran {\tt tclean} with `auto-multithresh' masking (with a noise threshold of 4.0, a sidelobe threshold of 2.0, a low noise threshold of 1.5, and a minimum beam fraction of 0.3), in interactive, multi-frequency synthesis mode, with a maximum number of cycle iterations of 10,000.
For data with no sources detected by auto-multithresh, {\tt tclean} produced dirty images, whereas images with auto-detected sources were cleaned to a depth of $2{\times}$ the {\tt tclean} estimated theoretical sensitivity.
Figure~\ref{fig:contgallery} presents a gallery of images, showing only the combined image for sources with multiple observations and excluding tracks that showed variable emission (see \S~\ref{sec:mm_var}).
Given the differences in time-on-source, antenna availability, and variations in the precipitable water vapor (PWV), the observations have a range of root-mean-square (RMS) sensitivities from 0.2--1.2\,mJy\,beam$^{-1}$, with mean and standard deviation 0.5\,mJy\,beam$^{-1}$ and 0.26\,mJy\,beam$^{-1}$ respectively.

We utilised the image domain fitting tool in {\tt CASA}, {\tt imfit}, to measure the flux of each detected source that we attribute to circumstellar emission.  These fluxes are presented in Table~\ref{tab:mainTable}.
Images with no clear emission were assigned upper limits on the target fluxes, defined as simply $3\sigma_{im}$, where $\sigma_{im}$ is the image-plane measured RMS.
Whilst we present primary-beam uncorrected images in Figure~\ref{fig:contgallery}, all fluxes were obtained from their corresponding primary-beam corrected images.
We corroborated all detections in the visibility domain by fitting measurement sets with the {\tt CASA} tool {\tt uvmodelfit}, fixing phase centres for each target, and the imaged locations for background sources.
Although five class~I and II YSOs were detected, the vast majority of class~III target observations presented no significant emission at their phase centers.
We detected significant millimeter emission associated with the class~III YSOs HD~28867, HD~283572, HD~31305, and HP~Tau G2, which we attribute to either gyro-synchrotron emission from stellar activity or emission from an unresolved binary companion (see \S\ref{sec:mm_var} and \S\ref{sec:obs_binaryComp}).

A stacked (median-filtered) image is presented in the bottom-right panel of Figure~\ref{fig:contgallery}, which combines only the images in the gallery without detections.
We detect no significant emission at the phase center of this stacked image, and therefore place a much deeper 3$\sigma$ upper limit on the median emission towards the class~III sample of $195\,\mu$Jy, with a $\sigma_{\rm im}=65\mu$Jy\,beam$^{-1}$.
This stacked non-detection is in line with expectations based on previous dedicated millimeter surveys of class~III YSOs \citep[i.e., as implied by the sensitivity and detection fraction in the Lupus class~III YSO survey, see][]{Lovell21a}.
To aid comparisons that we make later, we measured the median upper limit towards the class~III sample with stellar SpTs from B9--K3 as $290\,\mu$Jy.

\subsubsection{Variable millimeter emission}
\label{sec:mm_var}
Most survey targets were observed only once with the SMA, however 21 targets were observed in two-to-five independent tracks, and all sources were scanned ${>}3$ times in any given track that it was observed.
These multi-scan and multi-track observations provided an efficient check for variable (non-thermal) stellar emission that can masquerade as disk emission, in particular for class~III YSOs \citep[as seen for HD~283572 in the form of gyro-synchrotron emission, see][]{Lovell+2024}.

In the SMA sample, we found evidence that three class~III sources were variable: HD~283572 \citep{Lovell+2024}, HD~28867, and HP~Tau~G2.
For HD~28867 and HD~283572, this variability manifested in detections in individual tracks that were not present in subsequent tracks, i.e. variability on timescales of days--weeks.
To determine limits on the fluxes of targets in the HD~28867 and HD~283572 data, we only combined visibilities from data in which only quiescent levels of flux were observed.
For HP~Tau~G2, the variability manifested in strong intra-track flux alterations, on timescales of several minutes.
We confirmed this variability as physical (and not a calibration error) by comparing the fitted fluxes of HP~Tau~G2 and HP~Tau (a nearby bright disk, fortuitously located within the same field of view) in both the image and visibility domain on a per-scan basis.
We found that whilst HP~Tau had a consistent flux through the entire track ($35{\pm}2$\,mJy), HP~Tau~G2 by comparison varied from 0--30\,mJy over the same interval.
This finding may be unsurprising given the known radio brightness and variability of HP~Tau~G2 \citep[see e.g.][]{Cohen+1986, Rahman+2025}.

\subsubsection{Unresolved binary emission} \label{sec:obs_binaryComp}
HD~31305 was recently observed by ALMA and interpreted as a binary with a separation of 0\farcs580. 
In that work, \citet{Moor+2025} noted that the bright A-type primary component was not detected, but the lower-mass (K-type) class~II secondary has a flux consistent with that obtained in these SMA data.
We compared the best-fit emission location for HD~31305 with SMA data and found a consistent offset to that determined with ALMA (${\approx}0\farcs6$) and therefore also attribute the SMA detection to the class~II source.
Although the full K+21 sample does not include HD~31305~B, we renamed HD~31305 to HD~31305~A, updated its binary flag to `U' (unresolved), and in Table~\ref{tab:mainTable} report the flux of HD~31305~A as an upper limit based on the noise properties of the data, and add the detection of HD~31305~B to the lower section of the table.

\subsubsection{Archival SMA and ALMA data}
We collated available millimeter fluxes of the 512 sample sources from four studies with previous SMA and ALMA observations.
Class~0--II YSOs fluxes were obtained from \citet{Andrews13} and \citet{Akeson+2019}, and class~III YSO fluxes were obtained from \citet{Andrews13}, \citet{Hardy15}, \citet{Akeson+2019} and \citet{Michel21}.
We included source fluxes where these had been observed by either the SMA or ALMA, or had been observed by both, prioritising the SMA fluxes where possible, via the following logic: (1) if a source was observed only by either the SMA or ALMA, then we utilised the published fluxes or $3\sigma$ upper limits; (2) if a source was detected by both SMA and ALMA, we utilised the flux derived by the SMA; (3) if a source was detected by ALMA but not SMA, we utilised the flux derived from ALMA; and (4) if a source was not detected by either the SMA or ALMA, then we utilised the SMA derived upper limit.
There were zero instances in which source data was available for both instruments in which the SMA had detected a source that ALMA had not.
We prioritised data collation in this manner due to the much larger number of SMA class~III observations in comparison to ALMA, and so aimed to minimise potential systematic differences in the observations/flux calibration obtained from these two instruments' data.

Two Taurus sources from \citet{Michel21} are noted as detections, and these comprise the only class~III YSOs with disk detections in the full sample; V819~Tau, and XEST 08-033 (2MASS~J04354203+2252226).
In both cases, these sources' SEDs showed these systems to have mid- and/or far-infrared excess emission, with their millimeter fluxes broadly matching the expected extrapolation above their stellar photospheres (see \S\ref{sec:SED_data}), and thus we infer these ALMA detections to be from circumstellar dust.
Further ALMA data was available in Band~7 ($\approx$340\,GHz).
We imported the original Band~7 data in {\tt sdb} for SED modelling.
To include these data in the assessment of the luminosity distribution, we re-scaled these ALMA Band~7 (${\sim}0.89$mm) fluxes to 1.3\,mm fluxes assuming a submillimeter spectral slope of ${-3}$, i.e., $F_{\rm 1.3mm}{=}F_{\rm B7}(1.3/\lambda_{\rm B7})^{-3}$, in line with expectations that these would arise from optically thin emission, consistent with debris disks \citep[see e.g.][]{MacGregor16, Marshall+2017}.
Including the newly obtained SMA data, the full sample compiles 252 millimeter fluxes, 21 from class~0--F YSOs, 149 from class~II YSOs, and 82 from class~III YSOs, that we use to derive flux distributions and compare source properties in \S\ref{sec:results}.

Although a handful of additional targets were observed by the James Clark Maxwell Telescope (JCMT) and data were available from \citet{Andrews05}, these data are typically an order of magnitude less sensitive (extrapolated to 1.3\,mm) and are much lower resolution. Since those sources have less stringent upper limits, we did not include those data.

\begin{figure}
    \centering
    \includegraphics[clip, trim={0.0cm 0.0cm 0.0cm 0.0cm}, width=1.0\linewidth]{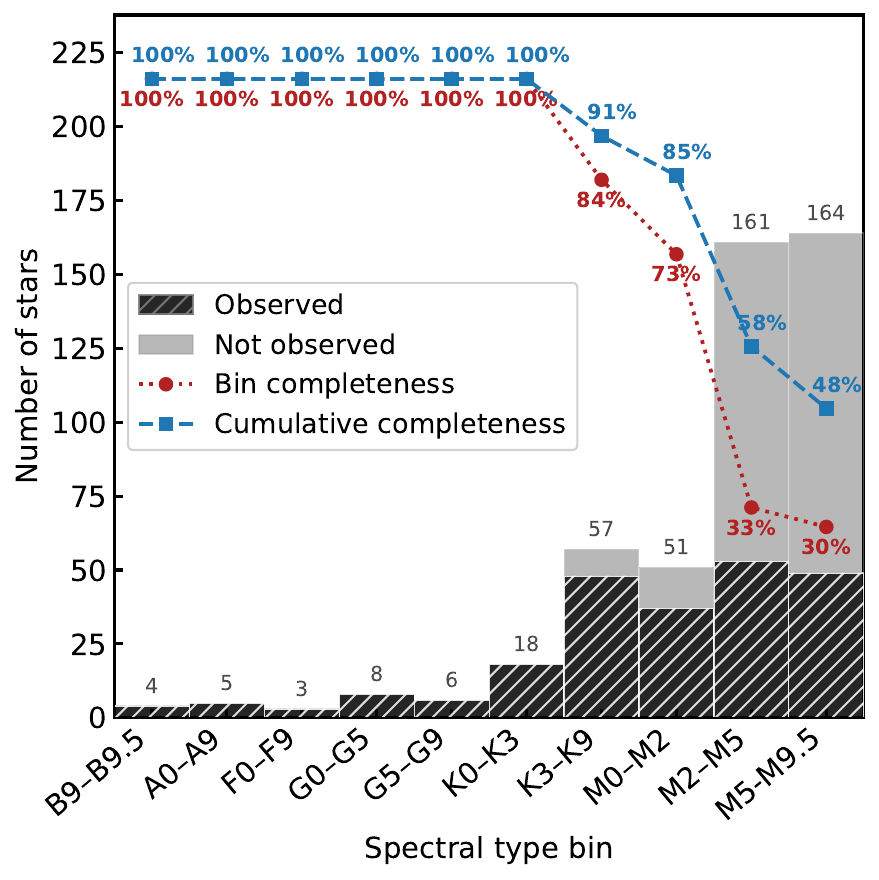}
    \caption{Histogram of class~II and class~III millimeter observations as a function of SpT. Completeness fractions are overplotted per-bin (red), and cumulatively for the sample (blue). The total of 477 excludes the 32 class~0--FS YSOs, and 3 YSOs without SpTs.}
    \label{fig:spt_completeness}
\end{figure}

\begin{figure*}
    \centering
    \includegraphics[clip, trim={3.2cm 1.8cm 3.5cm 2.75cm}, width=1.0\linewidth]{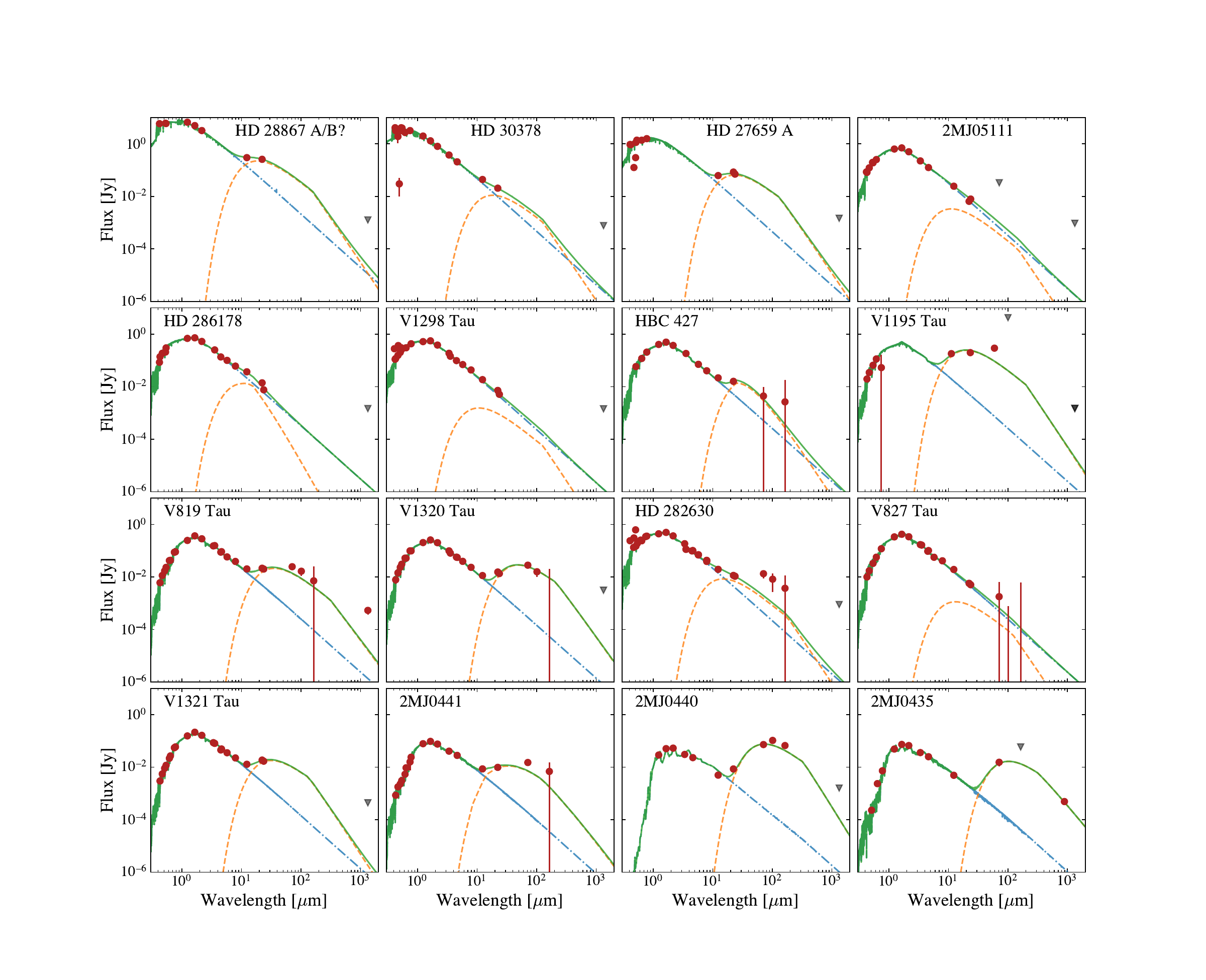}
    \caption{Gallery of 16 sources with significant thermal excesses. Blue, amber, and green lines indicate the best-fit stellar atmosphere model, disk model, and combined model.  Red points show the source photometry and grey triangles indicate photometric upper limits.}
    \label{fig:SEDs}
\end{figure*}

\subsubsection{Sample completeness}
Prior to this study, the largest collated dataset of (sub)millimeter interferometric class~III YSOs was 85 (total) in \citet{Michel21}, of which 12 class~III YSOs were members of Taurus.
So we have nearly doubled the total published census of millimeter data for class~III YSOs, and increased the Taurus sample by over a factor of five.
Although overall the YSO observations at millimeter wavelengths in Taurus remain incomplete, Figure~\ref{fig:spt_completeness} shows that the population of early SpT class~II and III YSOs have (nearly) all been surveyed to depths that can provide important constraints.
For example, Taurus class~II and III YSOs with SpTs earlier than K3 are 100\%\ observed (44/44), and earlier than M2 are 85\%\ observed (129/152), as shown in the blue cumulative curve on Fig.~\ref{fig:spt_completeness}.
This is sufficiently complete to construct millimeter luminosity distributions that are unbiased for YSO-class in Taurus, mitigating the survivor bias noted in \S 1 (see \S\ref{sec:results}).
Later-type class~II and III YSOs with SpT from M2--M9.5 remain 31\% observed (102/325) and thus we exclude these from later analyses.

\subsection{Multi-wavelength observations}
\label{sec:SED_data}
We imported all the collated millimeter data into {\tt sdb} to aid the analysis of the sample disk properties.
Before considering a YSO as host to a thermal circumstellar dust excess, we compared the best-fit {\tt sdb} models with and without extinction ($A_V$).

We examined the SEDs of the class~III YSOs and found 27 hosted significant thermal excess components from warm, faint emission, and cold emission that peaks well beyond $24\,\mu$m.
We report the model properties of all 27 excess sources in Table~\ref{tab:SEDVals} for completeness, noting that a handful are around mid-M stars, and that 11/27 have hot excesses that may instead be due to intrinsic stellar variability that we cannot rule out.
In total we detected significant excesses above stellar photospheric models towards 2${\times}$B, 1${\times}$A, 1${\times}$F, 1${\times}$G, 15${\times}$K, and 7${\times}$M-stars, with best-fit (single component) dust temperatures ranging from 44--978\,K.
All YSOs can be well-fitted with models comprising a stellar component and a single disk component.
We present the continuum spectra for the 16 targets with warm/cool excesses in Fig.~\ref{fig:SEDs}, with associated best-fit stellar and disk model components. 
A handful of SEDs are equally well-fit by a single star and two disk components (e.g., V819~Tau, {HD~282630, and 2MASS~J04210934+2750368). New high-angular resolution observations are necessary to determine whether these systems host multiple radially separated disk components, a radially extended dust distribution, or a single narrow disk, as the current data do not strongly favour any of these scenarios.
Both V1195~Tau~A and HD~28867~A have {\tt sdb}-derived excess emission fitted with a single disk component; however since we cannot confidently attribute this to either individual binary component A or B (or a combination), we report these as being from `A/B?' in Table~\ref{tab:SEDVals}.
We discuss these results further in \S\ref{sec:disc_seds}.

\section{Results} \label{sec:results}

\begin{figure}
    \centering
    \includegraphics[clip, trim={0.03cm 0.2cm 0.0cm 1.4cm}, width=0.99\linewidth]{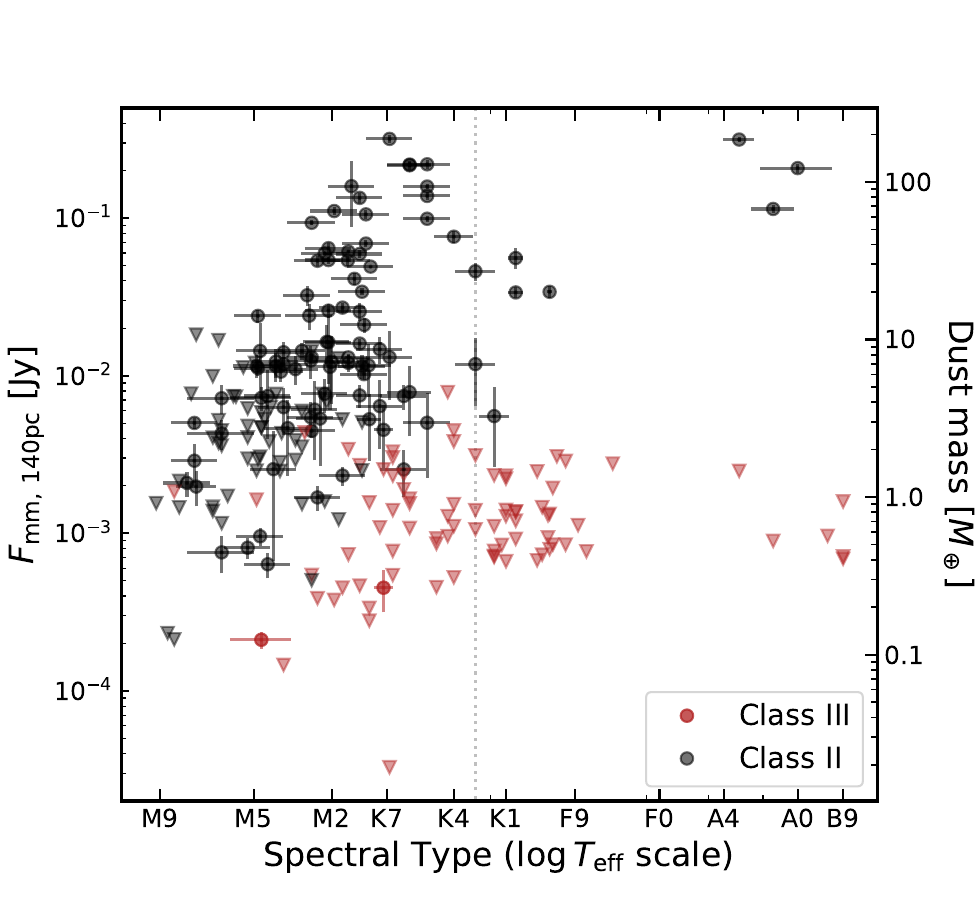}
    \caption{The millimeter luminosity distribution of Taurus YSOs as a function of stellar spectral type (presented on a log-$T_{\rm eff}$ scale), with corresponding (optically thin) dust masses presented on the right y-axis.  The class~III YSOs are marked in red, and class~0--II data are shown in black. We show a vertical bar at SpT = K3, earlier than which this survey is complete.}
    \label{fig:mmLumDist}
\end{figure}

\subsection{The Taurus YSO millimeter luminosity distribution}
Approximately one-third (82/252) of the millimeter-observed YSOs are at the class~III stage, of which just two were detected with millimeter circumstellar emission; V819~Tau, and XEST 08-033 (2MASS~J04354203+2252226).
In comparison, class~II (and earlier) Taurus YSOs yield 94 detections from 149 observations, despite having typically less sensitive millimeter (SMA) data.
The remaining class~0--FS sources in the sample yield 17 detections from 21 observations, though are not analysed further in this work.
The luminosities of both class~III detections are below all the luminosities of class~II detections irrespective.
Moreover, for all YSOs with an SpT earlier than K3 -- the SpT range for which this survey is complete -- all class~II YSOs were detected with luminosities exceeding all the upper-limit luminosities of class~III YSOs.
These findings clearly demonstrate the intrinsic millimeter faintness of class~III Taurus YSOs compared to their less-evolved counterparts.

Figure~\ref{fig:mmLumDist} presents the distribution of millimeter luminosities for Taurus class~II and III YSOs as a function of the stellar host spectral type.
This same information is also shown in the form of a left-censored Kaplan-Meier distribution in Figure~\ref{fig:KaplanMeier}, presenting the cumulative statistics for the 140\,pc--scaled normalised flux ($F_{\rm{mm,\,140pc}}$) binned into two SpT ranges (earlier than K3; and K3--M2). 
We made these plots with the statistical package {\tt lifelines} \citep{lifelines} which implements a Kaplan-Meier estimator as per \citet{KM_1958}.
In these bins, there are observed fractions of 42/42 and 85/108 respectively, or 36/36 and 56/70 respectively if known binaries/multiples are excluded.
Although not presented, the M2--M9.5 sources have much lower observed fractions of 102/325, and 70/265 excluding binaries.
We adopt $F_{\rm{mm,\,140pc}}$, which is directly proportional to the millimeter luminosity and defined as
\begin{equation}
    F_{\rm{mm,\,140pc}} = F_\nu \Big(\frac{d}{140}\Big)^2,
\end{equation}
for sources at a distance $d$ (parsec units) with millimeter fluxes of $F_\nu$ (Jy units).
We scale these fluxes to 140\,pc given this value is very close to the median distance to the Taurus sample of $142.5$\,pc.

\begin{figure}
    \centering
    \includegraphics[clip, trim={0cm 0.0cm 0.0cm 0.0cm}, width=0.99\linewidth]{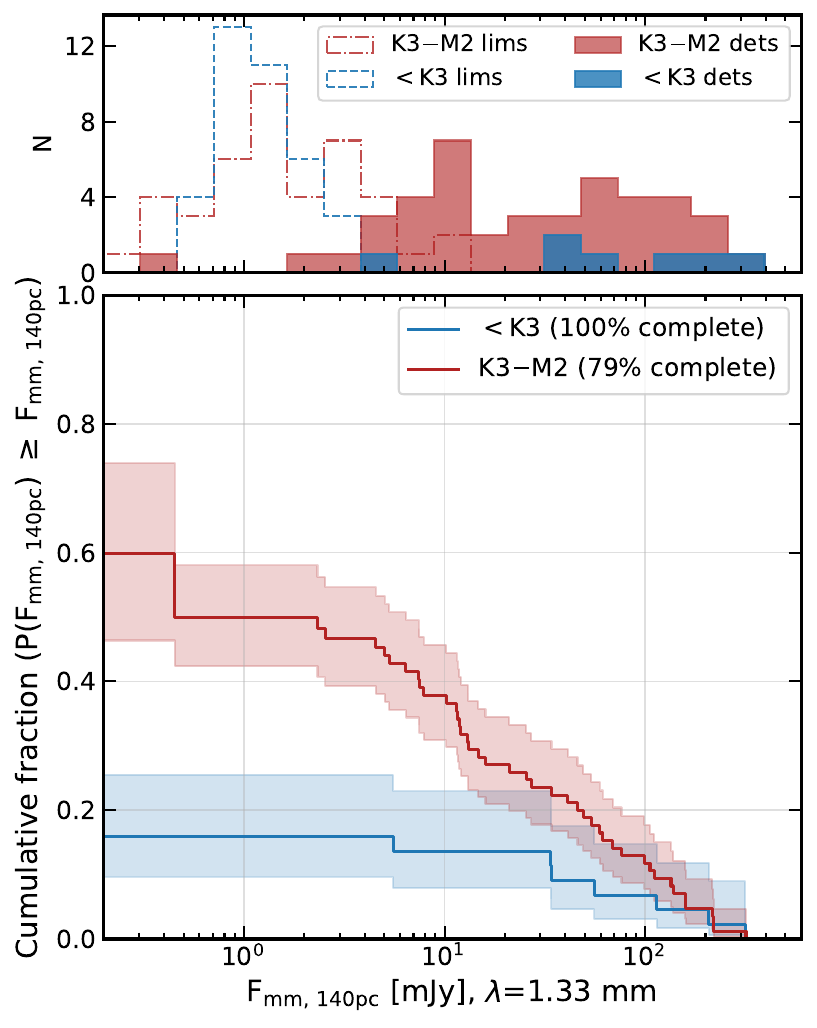}
    \caption{Cumulative distributions of millimeter luminosities for the combined sample of class~II and III YSOs, split into early- (blue) and mid- (red) SpT populations. The top panel shows histograms (binned differential distributions) of the two populations. }
    \label{fig:KaplanMeier}
\end{figure}

Figure~\ref{fig:mmLumDist} demonstrates that the scaled fluxes of all class~II YSOs with host SpT earlier than K3 (dotted line) are ${\approx}0.2$--2 orders of magnitude higher than the upper limits of all class~IIIs, where the separation between the class~III upper limits and class~II fluxes increases for earlier type stars (i.e. the scaled fluxes of class IIs are all ${>}1$ order of magnitude brighter than class~IIIs earlier than SpT K1, rising to 2 orders of magnitude for the A and B stars).
This can also be seen in the blue curve in Figure~\ref{fig:KaplanMeier}, which flattens from ${\approx}10$\,mJy down to the upper limits of the SMA data, demonstrating the lack of any millimeter bright YSOs in this flux range which we refer to as a break in the scaled fluxes.
Most interestingly, this break maps directly to YSO class, with all class~IIs above this break, and all class~IIIs below it.
The sample completeness in this SpT range demonstrates that this break is a real feature of Taurus YSOs, and confirms findings from previous (sub)millimeter surveys \citep[see e.g.][]{Duvert+2000, Andrews05, Hardy15, Lovell21a}.
In contrast, despite being much less complete, the late-type YSO distribution includes many detections that span this flux range.
The lack of an observed break in these later-SpT YSOs may reflect the combination of limited sensitivity to the class~IIIs and the intrinsically fainter class~II YSO millimeter emission for SpTs later than K3. 
Although the data cannot demonstrate a luminosity break for stars in the SpT range M2--K1 the class~II and III populations (48 class~II and 49 class~III observations) nevertheless remain statistically distinct in this range. 
Using a Gehan-type generalized Wilcoxon test with a permutation null, we find at least 86\% of class~II-class~III pairs have the class~II source brighter, none are ordered the other way, and the remaining 14\% of pairs comprise of the pairs where class~II non-detections cannot be compared with the class~III limits. 
The test reports $p<5\times10^{-5}$ (the effective resolution of the test for n=20,000 trials). The data therefore show that the luminosity distributions of the class~II and III populations over M2--K1 are distinct, even though there is no demonstrable luminosity break.
Overall, these data are a substantial improvement in the segregation of fluxes shown by previous-generation (JCMT-based) millimeter imaging, where the upper limit 1.3\,mm fluxes of class~III YSOs in Taurus were ${\gtrsim}10\times$ higher \citep{Andrews05}.

To understand if these distributions are significantly impacted by stellar multiplicity, we constructed the same figure excluding all sources with binarity flags.
We found the two distributions were impacted similarly, and thus that there appears to be no overall impact on the presence of a luminosity break if binaries are included or excluded.
It is possible that this finding reflects the low fraction of stellar multiples in the sample, or instead if this reflects the incompleteness of known stellar multiples.

\subsection{The Taurus YSO stellar mass and age distributions} \label{sec:stars}
As a corollary of fitting SED models, {\tt sdb} provides best-fit stellar luminosities and temperatures, with associated uncertainties.
Whilst we deemed these properties unreliable for class~II (and earlier) YSOs, these stellar parameters are all available for the earliest-type class~II YSOs from \citet{Andrews13}. 
Since the millimeter observations are largely complete for YSOs with spectral types earlier than M2, we determined the ages and stellar masses of all these early-type class~II and III YSOs using the publicly available code {\tt ysoisochrones} \citep{Deng+2026}, adopting the stellar models of both \citet{Siess00} (herein {\tt Siess}) and {\tt PARSEC} v2.0 \citet{Nguyen+2025}.
We excluded 4 YSOs from these fits owing to having poorly-modelled SEDs; these include V773~Tau (a variable class~II stellar quadruple), and the class~III YSOs HD~28867~B, V1195~Tau~B (discussed previously), and the class~III YSO UX~Tau~B \citep[also a double, and part of a stellar quadruple, see][]{Menard+2020}, with an SED partially confused with the class~II UX~Tau~A.
In total, we fit stellar masses and ages to 148 YSOs, i.e. 55 class~IIs, and 93 class~IIIs, which broadly reflects the class~II-to-class~III ratio in the star-forming region as a whole.
The class~II and III YSO stellar ages, masses and luminosities are shown as histograms in Figure~\ref{fig:stellar_properties} for the {\tt PARSEC} models, and we tabulate statistical properties for both models in Table~\ref{tab:stellar_properties}.

\begin{figure}
    \centering
    \includegraphics[clip, trim={0.1cm 0.1cm 0.2cm 0.1cm}, width=0.99\linewidth]{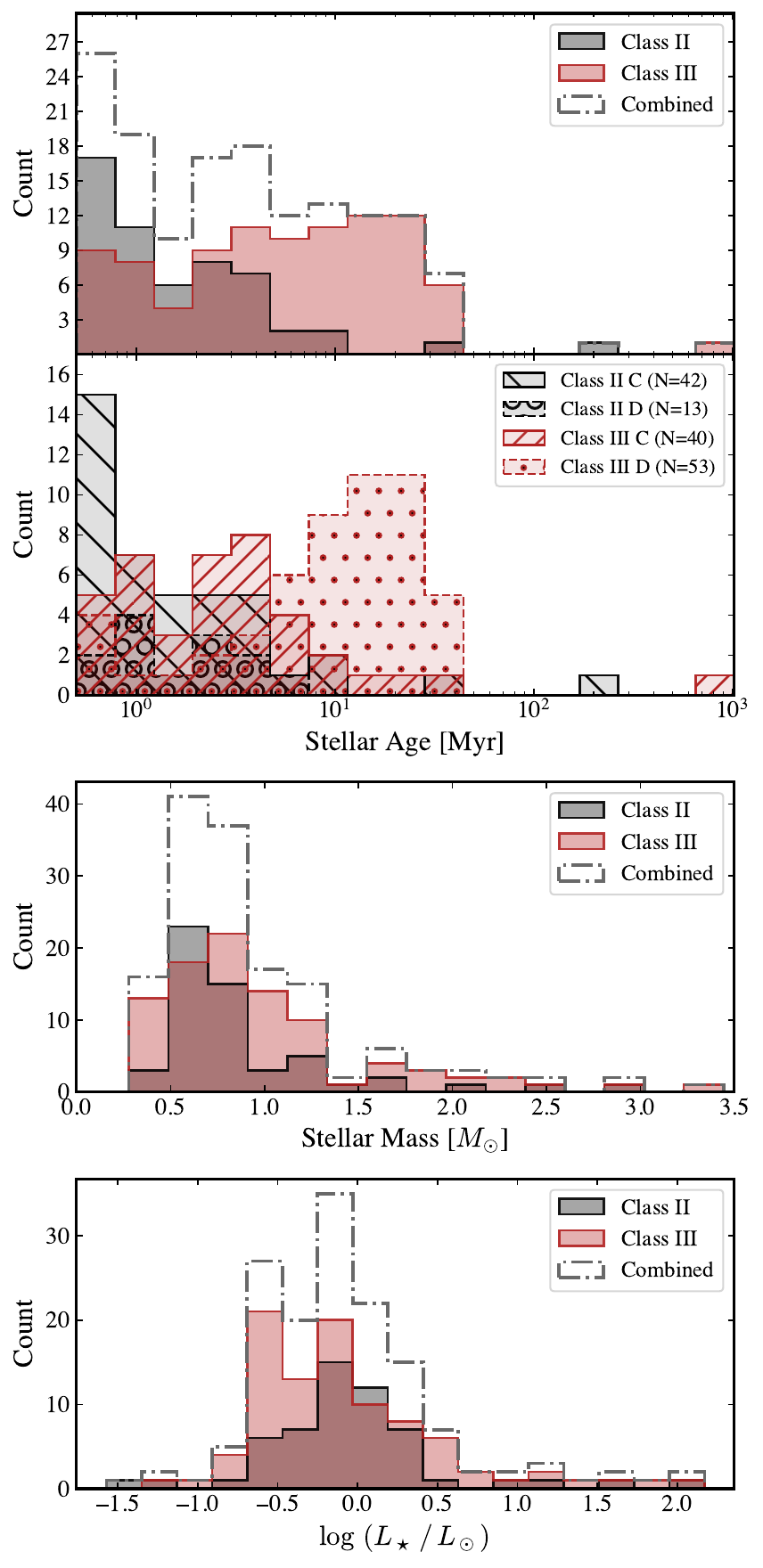}
    \caption{Histograms of stellar ages and masses fitted with {\tt PARSEC} stellar models, and stellar luminosities from SED analysis, for the earliest-type YSOs in the sample (148/152). }
    \label{fig:stellar_properties}
\end{figure}

\begin{table}[]
    \caption{Class~II and III stellar property statistics. We show the medians, and the means (and standard deviations) of the stellar mass and age distributions.}
    \centering
    \begin{tabular}{l | c c | c c}
    \hline
    \hline
    Sample & Med. Age & $\overline{ \rm{\log{t} }}$ & Med. $M_\star$ & $\overline{M_\star}{\pm}\sigma_{\overline{\rm{M_\star}}}$ \\
    & [Myr] & ${\pm}\sigma_{\rm{\overline{log{t}}}}$ & [$M_\odot$] & [$M_\odot$] \\
    \hline
    {\tt PARSEC} & & & & \\
    Class~II  & 1.2 & $6.2{\pm}0.5$ & 0.7 & $0.9{\pm}0.5$ \\
    Class~III & 5.7 & $6.7{\pm}0.6$ & 0.8 & $1.1{\pm}0.9$ \\
    \hline
    {\tt Siess} & & & & \\
    Class~II  & 2.4 & $6.4{\pm}0.3$ & 0.8 & $0.9{\pm}0.6$ \\
    Class~III & 7.9 & $6.9{\pm}0.6$ & 0.9 & $1.1{\pm}1.1$ \\
    \hline
    \end{tabular}
    \label{tab:stellar_properties}
\end{table}

We present statistical metrics describing the stellar age and mass properties of these two YSO samples in Table~\ref{tab:stellar_properties}, demonstrating that the medians and means of the class~III ages are larger than the class~IIs, whereas the stellar mass distributions overlap.
We performed two-sample Kolmogorov-Smirnov (K-S) and Anderson-Darling (A-D) tests, as well as two-sided Mann-Whitney (M-W) tests using the {\tt SciPy.stats} module \citep{SciPy}, on the distributions of stellar masses and ages to compare the class~II and III distributions for both model setups.
These tests confirm that the mass distributions are statistically consistent with the same parent population for both stellar model setups, with a null-hypothesis p-value of $p = 0.087$ for the K-S test, $p=0.21$ for the A-D test, and $p=0.36$ for the M-W test ({\tt Siess}), and a null-hypothesis p-value of $p = 0.005$ for the K-S test, $p=0.02$ for the A-D test, and $p=0.09$ for the M-W test ({\tt PARSEC}).
As expected from a visual inspection of the distributions in Figure~\ref{fig:stellar_properties}, the age distributions are inconsistent: we reject the null-hypothesis (identical parent age populations) with p-values $p=6\times10^{-9}$ ($\approx 5.8\sigma$ confidence) for the K-S test, $p{<}0.001$ for the A-D test\footnote{Note that the {\tt scipy} implementation of the A-D test has a p-value floor of 0.001. The A-D statistic for the two-age distributions has values of 18.7 ({\tt Siess}) and 18.8 ({\tt PARSEC}), both much larger than the critical value of 6.546 associated with $p=0.001$, and thus the evidence to reject the null hypothesis is even stronger given that the `true' p-value will be much less than 0.001.}, and $p=2\times10^{-7}$ for the M-W test ({\tt Siess}), and with p-values $p=7\times10^{-8}$ ($\approx 5.3\sigma$ confidence) for the K-S test, $p{<}0.001$ for the A-D test, and $p=9\times10^{-8}$ for the M-W test ({\tt PARSEC}).
The combination of these non-parametric statistical tests provides very strong evidence that there is a statistical difference in the distribution centers, tails, and medians of the class~II and III YSO age distributions.

Interestingly, there appears to be a significant spatial-dependence on the age demographics.
Of these 148 systems, K+21 designated 66 as members of Taurus' `distributed' sub-populations (`D'), with the rest as members of `core' sub-populations (`C', see second Stellar Age histogram), where these sub-population categories are spatially correlated, and which K+21 note have distinct age properties with D regions being typically older than C regions.
Using only the {\tt PARSEC} model values, if the distributed members are excluded from the population, then the median age of the class~II YSOs falls slightly from 1.2\,Myr to 1.1\,Myr, and the class~III median age falls substantially from 5.7\,Myr to 2.4\,Myr as evident comparing the different hatch types in the second Stellar Age panel in Fig.~\ref{fig:stellar_properties}.
The median difference between the two class populations is then just 1.3\,Myr, substantially less than 4.5\,Myr from the full sample in agreement with the findings of K+21.
This finding is corroborated by the {\tt Siess} models.
The class~II and III stellar mass distributions remain consistent between class~II and III YSOs after the exclusion of distributed members.

Overall, based on these stellar models, the class~III age distribution of this sample of YSOs is older than the class~II distribution, with the isochrone models suggesting a median difference of around 4--5\,Myr. 
In contrast, the stellar mass distributions of both YSO populations are consistent.
Taken together, these two distributions suggest that although the underlying class~II and III stars are drawn from the same population, these may have been born at different times.
This stellar analysis effectively reproduces the findings of \citet{Bertout+2007} where a similar isochronal analysis of stellar properties in Taurus showed a distinction in the age distribution of weak-line T-Tauri stars (WTTSs) and classical T-Tauri stars (CTTSs), with the WTTSs being inferred to be older.
It is perhaps unsurprising that we recover this finding, given the strong overlap in class~III YSOs with WTTSs and class~II YSOs with CTTSs, though we note this finding is largely driven by the inclusion of more spatially distributed YSOs.
This result provides important context for the analysis presented in both \S\ref{sec:dustmass} and \S\ref{sec:debrisDisks}.

\subsection{The Taurus YSO dust mass distribution} \label{sec:dustmass}
Directly from the millimeter fluxes, we can quantify disk \textit{dust} masses based on the expression
 \begin{equation}
     M_{\rm dust} = \frac{F_{\nu}d^2}{\kappa_{\nu}B_{\nu}(T)},
     \label{eq:mdust}
 \end{equation}
for sources with millimeter fluxes $F_{\nu}$, at distances $d$, with opacities $\kappa_{\nu}$, observed at frequency $\nu$  \citep{Hildebrand83, Beckwith00}, where we explicitly exclude gas from this estimate.
Since this expression is directly proportional to the millimeter luminosity, we can likewise infer the presence of a break in dust mass between class~II and class~III YSOs.

On the assumption that each source has only optically thin emission, a mean temperature $T{=}20$\,K \citep{Andrews05, Andrews20}, and a dust opacity of $2.25$\,cm$^{2}$\,g$^{-1}$ \citep[i.e., consistent with a $\beta=1$ spectral slope for dust with opacities of $10$\,cm$^{2}$\,g$^{-1}$ at 1000\,GHz, as per][]{Beckwith90}, we estimated the dust masses for all sources with millimeter data, shown on the right axis of Fig~\ref{fig:mmLumDist}.
Whilst these assumptions are crude and can underestimate the true dust masses for optically thick disks \citep[see e.g.,][]{Painter+2025}, these assumptions have been widely adopted by the community and thus serve as a good initial basis for comparison \citep[see e.g.][]{Andrews13, Pascucci16, Ansdell16, Ansdell18, Lovell21a}.
Nevertheless, given class~III YSOs will be optically thin at millimeter wavelengths, whereas class~II YSOs are more likely to be optically thick at millimeter wavelengths, this means that the dust mass break is more pronounced than the luminosity break (for fixed opacity and temperature), and the class~II dust masses are better interpreted as lower-limits.
Although this same dust mass break was shown to be present for YSOs with $M_\star{>}0.7\,M_\odot$ in Lupus by \citet{Lovell21a}, their sample was incomplete.  The finding we present here is significantly more robust, but taken together these results imply this break is a feature of (at least) early-SpT YSOs in general, and not just members of Taurus.

For the assumed optical properties and dust temperature outlined above, we find that the median dust masses of the early-, mid-, and late-type class~II YSO population are at least $33\,M_\oplus$, $15\,M_\oplus$, and $6.5\,M_\oplus$ respectively.
All of these values are lower-limits, given the unknown optical depths of class~II millimeter emission.
In contrast, the class~IIIs in those three bins have median dust masses ${<}0.70\,M_\oplus$, ${<}0.80\,M_\oplus$, and ${<}2.3\,M_\oplus$, respectively.
These large differences suggest that a significant amount (${>}99\%$ for the early-type YSOs) of dust mass has either been dispersed or transformed into larger solid bodies between the class~II and III stages.
Further, from the sample median upper limit from stacked data for the early type class~IIIs which has an associated dust mass limit of ${<}0.16\,M_\oplus$, the mass dispersal fraction increases to ${>}99.5\%$; around 3 orders of magnitude.

In \S\ref{sec:obs}, we presented a bifurcation in the $\alpha_{\rm IR}$ distribution, with few sources having any measurable infrared excess on the class~III side of the class~II--III boundary. 
This implied that warm dust is dispersed or transformed into larger solid bodies rapidly, given the large population of class~III YSOs with mid-infrared emission consistent with bare photospheres.
That we find a similar bifurcation in the millimeter data in terms of their fluxes/luminosities/dust masses implies that dispersal or solid body transformation acts on effectively the same timescale for the warm (inner disk) and cool (outer disk) dust, as found in other previous surveys of smaller samples \citep[see e.g.][]{Duvert+2000, Andrews05, Hardy15}.
Following the method of \citet{Koepferl+2013}, we estimate the timescale on which this dispersal or solid body transformation happens from $0.2{-}0.7$\,Myr assuming a disk evolution timescale of $\approx T_{\rm age}(N_{\rm trans}/N_{\rm tot})$, based on the fraction of disks transitioning between class~II and III stage, i.e. the number with ${-2.4}{<}\alpha_{\rm IR}{<}{-1.6}$ ($N_{\rm trans}$) divided by the population total ($N_{\rm tot}$) multiplied by either 1.3\,Myr, or 4.5\,Myr (the median age differences derived earlier). This timescale range broadly agrees with previous works derived from two-color estimates \citep[see e.g.][]{Ercolano+2011}. 
Considering these timescales as a proxy for the evolution from the class~II to III stage, the dust dispersal or dust to solid body transformation rate is, at least, {${\sim}$}50--165\,$M_\oplus\,$Myr$^{-1}$ (1.5--5${\times}10^{-9}\,M_\odot\,$yr$^{-1}$) for the earliest type stars.
Based on the crude assumption of a fixed disk gas-to-dust ratio of 100 \citep{Bohlin+78}, these results, based purely on the dust-continuum, agree well with complementary observations of disk gas. Observations of warm H$_2$ show that the upper limits on surface densities in the inner discs of class~III YSOs are ${<}0.01\times$ those of class~II YSOs \citep{Pascucci+2006, Ingleby+2009}, while upper limits on accretion in class~IIIs are also ${<}0.01-0.001\times$ the accretion rates measured for class~IIs \citep[][see also the discussion in \citealt{Alexander14}]{Alexander+2023, Thanathibodee+2023}. These findings may suggest that the dust and gas evolve on comparable timescales.

Theoretical models of disk evolution typically show that gas at large radii can be carried away in winds, either photoevaporative or magnetically-driven \citep[see e.g.][]{Clarke01, Alexander06, Bai16, Pascucci+2023, Bai2026}.
Less-well constrained is how, and where, dust at these radii can be so effectively removed, given that the maximum grain sizes of particles that can be entrained in winds is ${\approx}10\,\mu$m \citep{Takeuchi05, Hutchison+2021}.
Given that the millimeter data presented here, which are sensitive to grains in the hundreds of microns-centimeter size regime are cleared as rapidly as the micron-sized dust implies that another mechanism is required to clear the dust. 
Further, since the radial clearing and evolution of dust and gas appears to act on comparable timescales, it seems physically likely that whatever mechanism disperses the gas, also influences the dust, or provides conditions to allow the rapid evolution of the dust, i.e. not dispersal.
Two processes are likely in operation; these are grain removal and grain growth, with theoretical work suggesting that grain growth processes potentially dominate over grain removal processes \citep[with estimated conversion efficiencies of small grains to large bodies in the range 50--100\%, see e.g.][]{Simon+2024}.
The SMA data that we analysed in this work do not allow us to explore which grain removal processes dominates this stage of disk evolution, though this likely requires some combination of i) collisions and subsequent blow-out, and ii) inward radial drift and subsequent collisional fragmentation \citep[see e.g.][]{Birnstiel10, Owen19, Drazkowska23}.
We attempt in \S\ref{sec:debrisDisks} to measure whether these data argue in favour of dispersal or transformation being the dominant process that can interpret the large reduction in observed dust masses around class~III YSOs.

The presence of debris disks around main-sequence stars (and indeed the presence of planets) suggests that planetesimal formation is another mechanism plausibly in operation during protoplanetary disk dispersal \citep{Alexander07, Ying+2026}.
For example, as disks clear out gas and reduce the overall gas-to-dust ratio, this produces the conditions conducive to drive planetesimal formation via the streaming instability \citep{Youdin+2005, Johansen07, Carrera17, Carrera20, Ying+2026}.
The two class~III YSOs with millimeter detections in the Taurus sample have dust disk masses of $0.35\,M_\oplus$ and $0.09\,M_\oplus$, which are both below the maximum dust mass determined in the (SMA and ALMA) REASONS survey \citep[][]{Matra+2025}, meaning that the distribution of dust masses in Taurus' sampled class~III population overlaps with those of older debris disks.
This comparison may suggest that the dust detected around those two class~III YSOs arises from collisions or stochastic impacts within newly formed planetesimal disks, which are being observed as young debris disks \citep[as suggested for V819~Tau previously, see][]{Furlan+2009}.
It is nevertheless challenging to make more direct comparisons between the disk population of class~III YSOs and main sequence debris disks without resorting to population models.
We therefore consider the outcome of planetesimal belt formation in \S\ref{sec:debrisDisks} via population synthesis modelling to investigate how efficient planetesimal formation is between the class~II and III stages given the results outlined above, and whether there is substantial evidence for a population of young debris disks in Taurus.

\subsection{The disks of class~III YSOs in Taurus span a wide range of properties} \label{sec:disc_seds}
By analysing the multi-wavelength flux distributions of all 512 Taurus YSOs, we detected 27 with infrared emission excesses around class~III YSOs.
We associate 16 of these excesses with cool/warm circumstellar disk emission (${\approx}6\%$ of class~III YSOs), and 11 as possible hot circumstellar disk emission though given the possibility these may arise from stellar variability, do not interpret these further.
In the warm/cool excess sources, there is a broad continuum of fractional luminosities and blackbody radii inferred from the data, which indicates dust emission from scales close to the stellar photosphere out to tens of au \citep[i.e. given blackbody radii can underestimate physical radii by factors of a few][]{Pawellak14}.
The vast majority of excess sources are best explained as stars with disks, which exhibit fractional disk luminosities ($f = L_{\rm disk}/L_{\rm *}$) $f{<}0.01$, widely considered as the boundary between protoplanetary and debris disks \citep{Hughes18, Marino+2026}.
It is plausible that the larger $r_{\rm bb}$ class~III YSOs could form a population analogous to the debris disks observed on the main sequence (e.g., HD~28867A/B, HD~30378, HD~27659~A, and 2MASS~J04354203+2252226), whereas the smaller class~III YSOs with warm/hot excesses (e.g., HD~285281, HD~286178, HD~31281, and V1076~Tau) could have remnant primordial disks, or arise from small radii planetesimal belts that collisionally erode well-before the stellar main sequence. 
To be collisionally active, a planetesimal belt must be sufficiently stirred to maintain a collisional cascade. Although planetesimal belts may be born stirred, we estimated the self-stirring timescales of disks with radii ${\sim}10\,$au and planetesimal disk masses of $10\,M_\oplus$ to be ${\lesssim}1\,$Myr, adopting Equation~28 of \citet{Krivov18b}. In other words, disks with the implied $r_{\rm bb}$ values from SED fitting can be self-stirred on timescales shorter than the median stellar ages of the sample and thus be collisionally active.
The sample however shows no evidence for a correlation between stellar age and fitted $r_{bb}$. Whilst such a correlation might be expected in a population of collisional disks due to the more rapid grind-down of small radii disks \citep[e.g. as evident in][]{Matra+2025}, the lack of observed correlation in these data may simply reflect the small number of sources with $r_{\rm bb}$ measurements. 
We note that three (two cool and one warm) sources exceed the $f{=}0.01$ threshold by factors of 2.2--9.1 (including V1195~Tau, 2MJ0440, and 2MJ0421), and could arise from transition-disk-like objects/circumbinary rings (given that all of three of these systems are binaries).

As a simple way to consider whether any sources are indeed analogous to main sequence debris disks, we compared the distribution of derived $f$ and $r$ to the maximum fractional luminosity that a collisional planetesimal belt can have at a given age, i.e.
\begin{equation}
    f_{\rm max} = 0.58 \times10^{-9} r^{7/3} (dr/r) D_{\rm c}^{0.5} e^{-5/3} M_{\star}^{-5/6} L_{\star}^{-0.5} t_{\rm age}^{-1},
\end{equation}
as presented in Equation~18 of \citet{Wyatt08}, where we adopt constant values outlined in \S\ref{sec:debrisDisks_setup}.
We measured this distribution for the two median age differences between class~II and III YSOs (1.3\,Myr and 4.5\,Myr), and found that either 4 or 6 systems have fractional luminosities and radii consistent with being from a collisional belt. 
Specifically, the systems that meet this criteria are HD~30378, HD~28867~A/B, HD~27659~A, and 2MASS~J04354203+2252226 if $t_{\rm age}=4.5\,$Myr, and also HBC~427 and V1320~Tau if $t_{\rm age}=1.3\,$Myr.
We consider this subset of sources in more detail in \S\ref{sec:debris_comparison}.
Whilst these estimates do not rule out the possibility that the rest of the sample of cool/warm excesses are from collisional debris, this analysis suggests that only this subset can plausibly be from a steady-state collisional cascade.
For example, the 10--12 of the 16 cool/warm excesses may arise from remnant primordial dust, leftover from the protoplanetary disk, or from stochastic (rather than steady state) collisions, such as giant impacts and other terrestrial planet formation processes, expected to be in operation over the first ${\sim}$100\,Myr of evolution \citep[see e.g.][]{Kenyon04, Jackson+2012, Genda+2015}.

The SED analysis shows evidence of a trend seen on the main sequence for debris disks: despite M-stars being intrinsically more common, circumstellar dust excesses are more detectable around earlier-type stars \citep{Sibthorpe18, Lestrade+2025}.
We find that the earliest-type stars are intrinsically more likely to host thermal excess detections compared to later-type stars; for example considering only the warm/cool population, 50\% of the A and B stars have excesses, which falls to 9\% for the FGK-stars, and just 1\% for the M-stars.
We note that this may reflect a detection bias, given disks are intrinsically easier to detect around brighter stars following from the strong correlation between disk brightness and stellar SpT \citep[see e.g.][]{Ercolano+2009, Luppe+2020}.

\subsection{Robustness to sample selection} \label{disc:sampleSelection}
Whilst the initial sample of Taurus YSOs considered in this work were based on the 512 members identified by \citet{Krolikowski21}, we note that subsequent analysis in \citet{Luhman23a} determined a different population of YSOs in Taurus, totaling 532 members.
By cross-matching the {\it Gaia} IDs of both samples, we found 400 matches.
This suggests that the two samples are somewhat distinct, warranting a number of checks as to the robustness of our findings to the adopted initial sample.
This sample distinction largely results from \citet{Krolikowski21} requiring Gaia astrometry to be available (both parallax and proper motion), whereas \citet{Luhman23a} did not exclude such sources.
Given the 400 YSOs are simply a subset of the 512 from which we concluded earlier, we found that although the number of YSOs in the luminosity distribution of cross-matched YSOs was lower, the distribution maintained the same properties; i.e., a bifurcation in the millimeter luminosities between the class~II and class~III YSOs, and in the $\alpha_{\rm IR}$ distribution. 

By considering next the properties of the 132 YSOs in \citet{Luhman23a} that were excluded by \citet{Krolikowski21}, we found that just 1 would have been included in the early-SpT bin, the K0-type L1551/IRS~5, a class~0 YSO with a massive protoplanetary disk.
We therefore find that had we instead adopted the sample of \citet{Luhman23a}, our millimeter survey would remain 100\% complete for early-type YSOs, and thus would have left our conclusions based on these YSOs unchanged.
Of the remaining 131 YSOs, just 17 would have been included in the K3--M2 SpT bin, with the majority being well-studied class~I and II YSOs, including DP~Tau, HL~Tau, HH~30, HV~Tau~C, LR~1, IRAS~04016+2610, and V410~X-ray~2.
In this SpT range, the survey we presented would be $77\%$ complete (see Fig.~\ref{fig:KaplanMeier}) and given that a comparable fraction of these 17 YSOs have millimeter data (13/17), and that those with data are either detected with bright emission if class~II or not detected if class~III, then the more tentative conclusions that we drew from these data would remain unchanged for YSOs in this SpT range.
Given the low observation completion fraction of the later-type SpT range in the sample we adopted, we did not make any comparison to the remaining 118 YSOs in  \citet{Luhman23a}.
Overall, we find that the main conclusions drawn in this work would be the same irrespective of the origins of the initial sample selection.

\section{Debris disk formation} \label{sec:debrisDisks}
\subsection{Population synthesis model: planetesimal belts from class~II YSOs and their disks} \label{sec:debrisDisks_setup}
An important consequence of planetesimal formation is the development of planetesimal belts that can survive on to the stellar main sequence to become observable debris disks.
The mm continuum luminosities of nearby detected debris disks around main sequence stars are typically two orders of magnitude fainter than for class~II protoplanetary disks \citep{Wyatt15, Holland17, Lovell21a, Michel21}.
The survey presented in this study demonstrates that this large reduction in mm continuum luminosities from the class~II disks is already apparent at the class~III stage in Taurus, i.e., well before the stellar main sequence.
At the same time, class~III YSOs in Taurus still host faint but detectable levels of infrared emission from warmer disks, and in two cases, very faint millimeter emission from colder disks, as discussed in \S\ref{sec:disc_seds}.
These statistical constraints allow us to consider what happens at the class~II to III transition.

We constructed a population synthesis framework, based on the steady-state collisional evolution model of \citet{Wyatt07}.
In that model, the fractional luminosity of a planetesimal belt evolves as 
\begin{equation}
    f(t) = f(0) [1-t/t_c]^{-1},
    \label{eq:mtotEvol}
\end{equation}
for a belt that has collisionally evolved for a time $t$, with an initial fractional luminosity $f(0)$ at time $t{=}0$, and a collisional timescale $t_c$. 
The collisional timescale is defined as
\begin{equation}
    t_c = 1.4\times10^{-9}\frac{r^{13/3}(dr/r)D_{\rm c} Q_{\rm d}^{*5/6}}{M_{\rm tot}M_*^{4/3}e^{5/3}},
    \label{eq:time_coll}
\end{equation}
for a disk with mean radius $r$, fractional width $dr/r$, planetesimal strength $Q_D^*$ (fixed at 150\,J\,kg$^{-1}$), maximum planetesimal diameter $D_c$ (fixed at 60\,km), mean planetesimal eccentricity $e$ (fixed at 0.05, and assumed equal to the planetesimal inclination as-per the original steady-state collisional evolution model), and mass $M_{\rm tot}$.
The collision timescale expression shows that radially small, narrow, massive, and more collisionally active disks (here parametrised by planetesimal eccentricity), as well as disks around more massive stars, all grind down faster because of their shorter $t_c$.

The total planetesimal disk mass is directly proportional to the fractional luminosity,
\begin{equation}
    f = \frac{0.37 M_{\rm tot}}{r^2D_{\rm bl}^{0.5}{D_{\rm c}^{0.5}}},
    \label{eq:Mtot_f_r2}
\end{equation}
for disks with blowout size $D_{\rm bl}$, defined as {$D_{\rm bl} = 0.8(L_\star/M_\star)(2700/\rho)$}, where we fix the grain density as $\rho=2700$\,kg\,m$^{-3}$ \citep{Wyatt08}.
These model expressions (Equations~\ref{eq:time_coll} and \ref{eq:Mtot_f_r2}) result from assuming a single grain size distribution from the smallest grains up to the largest planetesimals that follow $n(D)\propto D^{2-3q}$ with $q=11/6$, which follows from \citet{Dohnanyi69}, and which implies that the disk mass is dominated by the largest bodies, whereas the total cross sectional area is dominated by the smallest bodies. 
We verified the functionality of the model by reproducing the older main sequence A-star simulation outcomes of \citet{Wyatt07}, using the same choice of stellar and disk properties outlined in that work.
Since we are simulating a steady-state collisional cascade, these models will not increase the mid-infrared and sub-millimeter disk fluxes, as predicted to occur over the first 10s of Myr by the collisional models of \citet{Kenyon+2008}. We do not account for stochastic collisions, but note that a more careful treatment may be necessary when analysing disks in stellar populations that are older than those in Taurus, e.g. such as in young moving groups, where it has been shown that fractional disk luminosities may drop much faster than predictions from steady-state models \citep[see e.g.][]{Pawellek+2021}. 

\subsection{Stellar and disk properties from the Taurus survey}
We synthesised a stellar population which follows a log-normal mass distribution with $\mu=0.7\,M_\odot$ and $\sigma=0.2$\,dex, which broadly matches the stellar distribution of Taurus stars in the B9--M2 range (see Fig.~\ref{fig:stellar_properties}).
We simulate each of these stars with a luminosity that follows a scaling relation $L_\star \propto M_\star^2$, which also broadly matches the measured luminosity distribution of the sample.
Whilst these stellar mass and luminosity distribution prescriptions are simple, by sampling steeper, shallower and broader mass and luminosity distributions, we found that these generally have little impact on the statistical properties of the output disk population.

We adopt an initial disk mass distribution, log-normally distributed with a standard deviation of 0.5 dex, and a mean scaling that follows the relationship between stellar mass and disk (dust) mass derived by \citet{Andrews13},
\begin{equation}
  M_{\rm disk} = \varepsilon_{\rm p} \times 10^{-2.3} M_{\star}^{1.4}.
\end{equation}
We modified this scaling relation with the inclusion of a planetesimal formation efficiency, $\varepsilon_{\rm p}$.
By sampling a range of $\varepsilon_{\rm p}$, this model allows us to investigate the formation of collisional planetesimal belts on the assumption that some fraction ($\varepsilon_{\rm p}$) of protoplanetary dust mass is converted into planetesimals.  The planetesimal belt mass ($M_{\rm tot}$) is equal to the initial debris disk mass in each simulation.
We adopted a disk radius distribution that follows a power-law distribution $n(r)\propto r^\gamma$, for $\gamma{=}{-0.25}$, between radii of 1--200\,au, on the broad assumption that ALMA observations of class~II disks typically sample the radii at which planetesimals are formed.
This radius distribution assumes that all protoplanetary disks form belts with radii drawn from this $n(r)$ distribution, with fixed fractional widths ($dr/r$), which we set to 0.8 to match the median value derived by the REASONS survey \citep[see][]{Matra+2025}.
This radius distribution produces a population that has ${\approx}25$\% of disks with radii larger than 25\,au \citep[which broadly matches the observed fractions of so-called `compact' and `structured'/large protoplanetary disks by ALMA, see e.g.][]{Najita+2022, Manara+2023}.
The exponent in this radius distribution is shallower than, but broadly consistent with that adopted in \citet{Wyatt07}, based on JCMT data for a relatively small sample of main sequence debris disks \citep{Najita+2005}.
The model does not couple disk masses and radii.

\begin{figure}
    \centering
    \includegraphics[clip, trim={0.1cm 0.3cm 1.2cm 1.3cm}, width=1.0\linewidth]{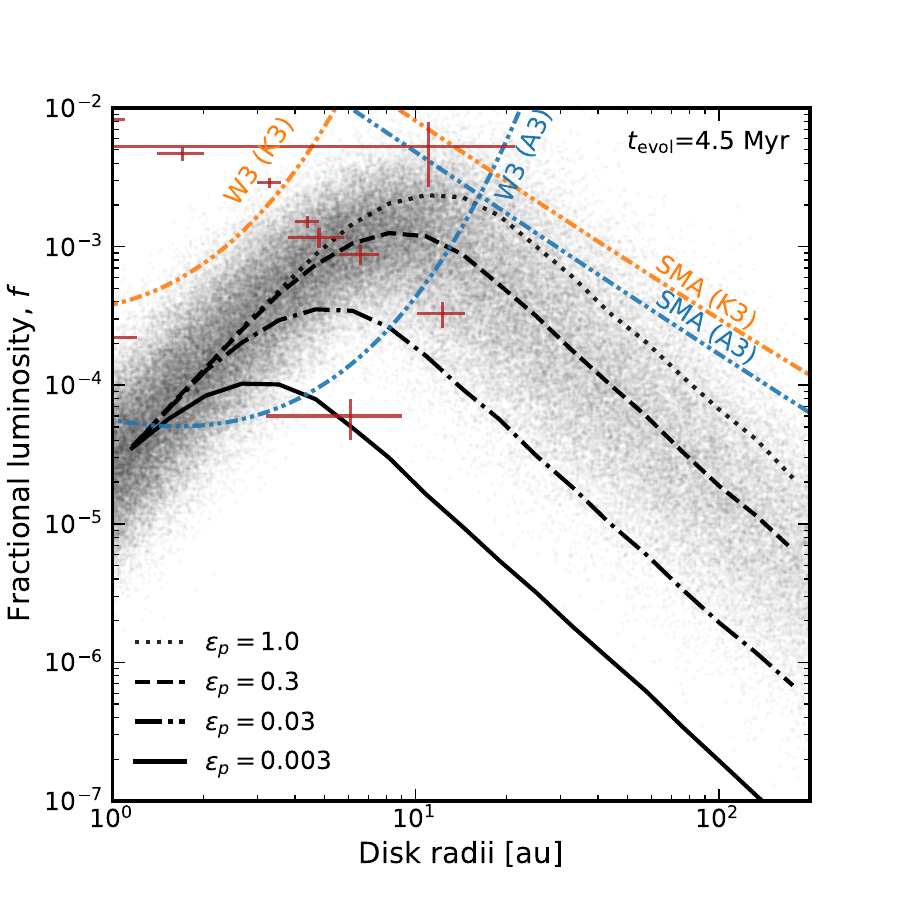}
    \caption{Fractional luminosity versus disk radius for the model population (100,000 sources per $\varepsilon_p$, with black dots shown for $\varepsilon_p=0.3$), which shows the collisional evolution of disks over 4.5\,Myr, illustrating the median fractional luminosity distributions for different planetesimal formation efficiencies with black lines. We over plot the detection thresholds for a K3 star (amber) and an A3 star (blue) with {\it WISE W3}, and SMA (1.3\,mm), and in red show the SED-fitted $f$ and $r_{\rm bb}$ values for class~III YSOs from Table~\ref{tab:SEDVals}.}
    \label{fig:beltmodel}
\end{figure}

\subsection{Modelling scenarios}
We sampled 100,000 realizations randomly from these star and disk distributions for six values of $\varepsilon_{\rm p}$, 0.0001, 0.0003, 0.003, 0.03, 0.3, and 1.0, and evolved each of these as per Eq.~\ref{eq:mtotEvol} for either $t=1.3$\,Myr and $t=4.5$\,Myr, the approximate difference between the median ages of the class~II and III YSOs (see Table~\ref{tab:stellar_properties}, i.e.  either excluding/including the distributed members).
We present in Fig.~\ref{fig:beltmodel} four of the modelled population median distributions (black lines, with distinct line-styles), and the simulation outputs for the $\varepsilon_{\rm p}{=}0.3$ setup (grey dots) for the 4.5\,Myr simulation.
The red data points plotted in Figure~\ref{fig:beltmodel} mark the SED-detected disks, and the detection thresholds for {\it WISE} $W3$ and the SMA (1.3\,mm) are marked with orange curves for a disk illuminated by either a K3 star (amber) or an A3 star (blue) at 140\,pc, which both show the effective instrumental sensitivities to the simulated Taurus populations.
The absolute calibration limit of {\it WISE W3} is 4.5\% \citep{Jarrett+2011}, therefore detecting excess disk emission above a stellar photosphere at the $3\sigma$ level practically requires the disk flux to exceed 13.5\% of the stellar flux, placing any {\it W3} detections in the `calibration-limited' regime \citep[where we adopt a calibration-limited uncertainty value $R_\nu{=} 0.135$, following the procedure of][]{Wyatt08}.
The millimeter sensitivity to disk emission is instead sensitivity limited, and presented as based on the $3\sigma$ median flux limit of 1.4\,mJy, as presented in \S\ref{sec:obsSMA}.
We note that the differing disk sensitivities of the SMA and \textit{WISE W3} are primarily driven by the disk radial distribution and their subsequent collisional evolution, from which we expect intrinsically larger detection rates with \textit{WISE W3} in comparison to the SMA.

As expected from the input disk radius distribution, in all cases the population is densely populated towards the left of the diagram, demonstrating the higher fraction of smaller disks in comparison to larger disks. 
And given the few-ten au turnover, these planetesimals collisionally grind down very fast, suggesting that many debris disks should be undetectable even after four Myr, with only the brighter end of the distribution clearly detectable with {\it WISE} data.
The simulations suggest that the largest disks, with radii exceeding 10--20\,au, retain their primordial fractional luminosities after four Myr (and hence follow the $f\propto r^{-2}$ behaviour implied by Eq.~\ref{eq:Mtot_f_r2}) with the brightest end of the distribution detectable with SMA data.
In general, these sensitivity curves suggest that the SMA and {\it WISE} data are largely insensitive to simulated disks in Taurus around early-K (and later) type stars, but that the combined mid-infrared and millimeter detection fractions (both for the full sample, and for different SpT ranges) can offer insights as to which setup ($\varepsilon_{\rm p}$ value) best reflects the data.

\begin{table}[]
    \caption{Comparison between the detection counts of sources with SpTs earlier than K3, with either {\it WISE W3} or the SMA, for the simulated populations at 1.3\,Myr (top) and 4.5\,Myr (bottom), alongside their comparative data.}
    \centering
    \begin{tabular}{l | c c | c}
    \hline
    \hline
    Scenario: $t=1.3\,$Myr & W3 & SMA  & Consistent?  \\
    \hline
    Data: $f_{\rm max:\,1.3\,Myr}$ & 3 & 0 & -- \\
    $\varepsilon_p=1.0$ & 6 & 3 & No \\
    $\varepsilon_p=0.3$ & 6 & 0 & Yes \\
    $\varepsilon_p=0.03$ & 4 & 0 & Yes \\
    $\varepsilon_p=0.003$ & 2 & 0 & Yes \\
    $\varepsilon_p=0.0003$ & 1 & 0 & Yes \\   
    $\varepsilon_p=0.0001$ & 0 & 0 & No \\   
    \hline
    \hline
    Scenario: $t=4.5\,$Myr & W3 & SMA  & Consistent?  \\
    \hline
    Data: $f_{\rm max:\,4.5\,Myr}$ & 3 & 0 & -- \\
    $\varepsilon_p=1.0$ & 3 & 3 & No \\
    $\varepsilon_p=0.3$ & 3 & 0 & Yes \\
    $\varepsilon_p=0.03$ & 2 & 0 & Yes \\
    $\varepsilon_p=0.003$ & 1 & 0 & Yes \\   
    $\varepsilon_p=0.0003$ & 0 & 0 & No \\   
    $\varepsilon_p=0.0001$ & 0 & 0 & No \\   
    \hline
    \end{tabular}
    \label{tab:sim_det_count}
\end{table}

\subsection{Comparison to data} \label{sec:debris_comparison}
Since the population synthesis model predicts the properties of collisional belts, here we compare the detection counts of YSOs that have best-fit $f$ and $r$ values consistent with being from a collisional cascade (see \S\ref{sec:disc_seds}) to the expected detection counts from each simulation.
These comparisons account for the observed sample size, and instrumental sensitivity curves, shown on Fig.~\ref{fig:beltmodel}.
For ease of comparison, we compare these counts to the single largest class~III SpT bin that has been observed in a complete sense, i.e. stars earlier than K3 (32 total).
In this SpT bin, the number of YSOs with SED excesses consistent with a collisional cascade evident from {\it WISE W3} is 3, and from the SMA is 0, irrespective of the assumed age/collisional timescale assumed (either 1.3\,Myr or 4.5\,Myr).
Since the predicted model counts alter when choosing either of these timescales, we compare simulations from both ages to the data.

We tabulate simulated counts from the twelve model scenarios and those from the data in Table~\ref{tab:sim_det_count}.
Comparing each data row with their corresponding simulations, we find that the $\varepsilon_{\rm p}{=}1.0$ and $\varepsilon_{\rm p}{=}0.0001{-}0.0003$ models are generally incompatible, given these either over-predict the number of millimeter detections for $\varepsilon_{\rm p}{=}1.0$, and under-predict the number of {\it WISE W3} excesses (by 3) for $\varepsilon_{\rm p}{=}0.0003$ and/or $\varepsilon_{\rm p}{=}0.0001$.
The intermediate scenarios $\varepsilon_{\rm p}{=}0.003{-}0.3$ generally offer a better match to the data, correctly predicting 0 SMA detections in all scenarios, and correctly predicting the {\it WISE W3} excess counts to within statistical uncertainties.
Although these comparisons are based on small number counts, these are derived from a population of 32 stars.
For example, given a mean empirical expectation value (detection frequency) of $3/32$ (9.4\%), the probability of observing 0 {\it WISE W3} excesses is 4.3\%.
We therefore rule out the $\varepsilon_{\rm p}{=}0.0001$ scenario at $p=0.043$ (95\% confidence), and the 4.5\,Myr $\varepsilon_{\rm p}{=}0.0003$ scenario at this same confidence.
Coincidentally, both $\varepsilon_{\rm p}{=}1.0$ scenarios predict 3/32 SMA detections, which compared to the 0/32 observed, results in the same probability of 4.3\% that these are consistent.
We therefore also rule out the $\varepsilon_{\rm p}{=}1.0$ scenario at $p=0.043$ (95\% confidence).

We finally note that the median flux of all simulated populations (over the full range of $\varepsilon_{\rm p}$) is below the median-stacked upper limit from the SMA data of sources earlier than K3 (\S\ref{sec:images_fluxes}). 
Since that upper limit is $\sim$5$\times$ deeper than the median upper limit used in the sensitivity threshold comparisons presented above, this comparison does not yield a significantly improved constraint on $\varepsilon_{\rm p}$.

\subsection{Planetesimal formation efficiencies from YSO surveys}
The simulated population suggests a broad plausible range of planetesimal formation efficiencies of a few-tenths to a few tens of percent, i.e., we cannot yet say whether or not during transition from class~II to III YSO stage if the bulk mass fraction of dust is dispersed rather than converted into planetesimals. 
More generally, we have shown how the combination of mid-infrared and millimeter photometry of class~II and III YSO populations can constrain planetesimal formation efficiency.
However, this suggestion comes with four core assumptions: (1) that the youngest planetesimal belts share the same planetesimal properties as those around mature debris disks; (2) that the amount of dust present at the class~II stage provides a reasonable proxy for the maximum available solid material to form planetesimals; (3) that planetesimals most typically form in belts; and (4) that no dust is converted into planets.
The first assumption is challenging to test with these data alone, given the degeneracies between e.g. $e$, $D_{\rm c}$ and $Q_{\rm d}^\star$ \citep[see discussion in][]{Wyatt07}, but seems a reasonable first-order estimate.
The second assumption implies that the formation efficiencies that we simulated are (at worst) upper limits, given that we neglected the possibility that planetesimals are already present in class~II disks \citep[as argued for by e.g.][]{Gerbig+2019, Bernabo_2022, Williams+2024}. 
For this modelling framework, accounting for this would effectively raise the initial disk mass, and thus the data would be more closely matched by lower values of $\varepsilon_{\rm p}$.
Alternatively, this could imply that all planetesimals are already present before the class~II to III transition, and do not require any primordial dust conversion.
This second assumption also mimics the very real nuance that the class~II dust masses are also lower limits, given their millimeter fluxes are optically thick, and their dust masses are estimated assuming these to be optically thin; this would also argue for lower values of $\varepsilon_{\rm p}$.
Relaxing the third assumption could imply that the overall efficiency with which planetesimals are formed is higher, but that only a subset of planetesimals form in belts and then subsequently produce collisional debris. 
As such, we note that the value of $\varepsilon_{\rm p}$ considered here is for now constrained within a relatively loose range, and that it may only define the planetesimal \textit{belt} formation efficiency.
Finally, since some fraction of dust may go on to form larger planet-sized bodies that do not contribute to the collisional cascade, this may argue for needing higher values of $\varepsilon_{\rm p}$.
Given these competing effects, further simulations to investigate how these assumptions influence planetesimal formation efficiencies are beyond the scope of this work, but are clearly now warranted.

Overall we therefore posit an important corollary: by sampling the youngest planetesimal belts that are yet to have reached collisional steady-state, such population models suggest one can constrain the efficiency with which planetesimal belts are formed between the class~II and III stages of YSO evolution, given the caveats outlined above.
We note that although there are no large class~III disks detected with the SMA, if this outer region can be sufficiently well-sampled (e.g., with more sensitive ALMA data), this model could provide far better quantitative constraints on the initial disk mass distribution, and planetesimal formation efficiency.
Given the distinct population demographics at different ages (and underlying differences in stellar and disk properties), these findings and parameter degeneracies can be more robustly determined by sampling different star-forming regions at different collisional ages.

\section{Conclusions} \label{sec:conclusions}
We presented an SMA 1.3\,mm (225\,GHz) continuum survey of YSOs in the 2--3\,Myr Taurus-Auriga star-forming region, based on the sample derived by \citet{Krolikowski21}. The key results of this work are:

\begin{enumerate}
  \item We observed 69 class~III and 5 class~I/II YSOs spanning spectral types M3--B9, and place these observations in the context of a larger millimeter sample totalling 149 class~II and 82 class~III YSOs that constitutes the largest millimeter survey of Taurus YSOs to date. This sample is observationally complete (100\%) for spectral types B9--K3 and substantially complete (85\%) from B9--M2 at millimeter wavelengths, with a median $3\sigma$ sensitivity of 1.4\,mJy.

  \item We measured the complete distribution of infrared spectral indices for the sample and found a clear bifurcation between class~II and III YSOs, suggesting the evolution between these stages to be fast given the lack of YSOs caught in the transitionary stage. We further quantified the fractional populations of YSO classes in three broad spectral-type bins, and showed that the class~III fraction peaks toward the earliest-type stars.  This suggests that disk evolution in Taurus is more advanced for higher host star masses.

  \item Although ALMA observations identified thermal dust emission from two class~III YSOs in this sample, we do not detect thermal 1.3\,mm emission attributable to circumstellar dust around class~III YSOs with any SMA data. We identified a small number of class~III detections that are not consistent with dust emission: three sources can be attributed to variable stellar activity, and one detection is consistent with unresolved thermal emission from a class~II companion. Together, these results indicate that detectable cold dust reservoirs are largely absent by the class~III stage.

  \item We constructed spectral energy distributions and fitted star and disk emission models for the sample, identifying 27 class~III YSOs that exhibit thermal excess emission in the infrared. These excesses demonstrate that, while millimeter-bright cold dust reservoirs are generally not present among class~III objects, a subset of sources retain detectable thermal emission at shorter wavelengths, arising from dust at a range of temperatures. 

  \item Using the stellar photosphere properties of the earlier-type stars, we find that the masses of the class~II and class~III YSO populations are consistent, but a significant fraction of the class~III YSOs are older than their class~II counterparts by $\sim4{-}5$\,Myr. This age difference is consistent with an evolutionary picture in which systems progress from class~II to class~III over ${\sim}4$--5\,Myr timescales. We however found a spatial dependence on this age difference, with the median difference falling to ${\sim}1.3$\,Myr if `distributed' members of Taurus were excluded from the stellar age analysis, suggesting the class~II to class~III transition may happen on more rapid timescales. We estimate the transition timescale from 0.2--0.7\,Myr based on these age ranges and the distribution of mid-infrared spectral slopes.

  \item Using all available Taurus millimeter measurements, we constructed continuum luminosity and dust-mass distributions, and present evidence for a bifurcation between the class~II and class~III populations. Interpreted in terms of dust evolution and using sample statistical averages, this bifurcation implies that ${\gtrsim}99.5\%$ of the dust is dispersed or transformed into larger solid bodies as YSOs transition from the class~II to class~III stage around earlier-type spectral classes. Due to sensitivity limits, these data cannot determine if there is a spectral-type dependence on the dispersal fraction, underlining the need to observe the mid- and late-type class~III YSOs with more sensitive millimeter observations. 

  \item The combined infrared and millimeter constraints indicate that warm (inner) and cold (outer) dust reservoirs evolve on comparable timescales; i.e., there is a rapid evolution across the entire radial extent of the protoplanetary disks. We discuss how the rapid dust evolution connects to the observed rapid gas evolution, and posit where and how dust is dispersed or transformed into larger solid bodies (planetesimals).

  \item We compared the observed infrared--millimeter properties of class~III YSOs with population synthesis models for collisional debris disks. We constructed a population of planetesimal belts, derived from the Taurus class~II disk and stellar properties, and compared models that assumed different dust-to-planetesimal formation efficiencies at two evolutionary times.
  The data are broadly consistent with a belt population that formed planetesimals from the dust disks with an efficiency of a few-tenths to a few tens of percent, and that is now evolving toward the main sequence as young debris disks. Whilst these constraints are not yet stringent, we highlight that multi-wavelength surveys comparable to this work undertaken in different star-forming regions can help to measure the conversion efficiency of dust to planetesimals.
\end{enumerate}

In combination, these results support a picture in which the transition from class~II to class~III in Taurus is accompanied by extremely efficient removal of millimeter-detectable dust, with the dust either being dispersed, or converted into planetesimals/planetesimal belts. A fraction of these systems emerge from the protoplanetary phase with planetesimal belts capable of sustaining detectable debris disk emission.

\section{Software and third party data repository citations} \label{sec:data}
This paper makes use of the following SMA data: 2021B-S014, 2023B-S038, and 2025B-S040 (PI: J. B. Lovell), 2021A-S070 and 2021B-S012 (PI: S. Andrews), and 2021B-H003 (PI: J. P. Williams). The specific SMA observation IDs utilised from these projects are as follows: 10871, 10872, 10910, 10943, 10950, 10962, 10991, 11000, 11002, 11003, 11006, 11007, 11010, 11012, 11019, 11021, 11025, 11028, 11045, 11051, 11076, 11078, 11105, 11128, 11135, 11137, 11142, 11145, 11151, 11159, 11178, 12491, and 14137.
For all 33 new SMA tracks associated with this publication we provide their image and measurement set, and their reduction scripts (see \S\ref{sec:obsSMA}) which we make publicly available on Zenodo, see \citet{Lovell_26_Taurus_data}. 

\facilities{Smithsonian Astrophysical Observatory (SAO)/Academia Sinica SubMillimeter Array (SMA) at Mauna Kea Observatory, Atacama Large Millimeter/subMillimeter Array (ALMA)}

\software{This research made use of the following software packages and tools:
          {\tt Astropy} \citep{astropy:2013, astropy:2018, astropy:2022},
          {\tt CASA} \citep{CASA07},
          {\tt COMPASS}, Keating et al. (in prep.),
          {\tt lifelines} \citep{lifelines}, 
          {\tt Matplotlib} \citep{Matplotlib},
          {\tt NumPy} \citep{NumPy}, 
          {\tt pandas}\citep{pandas}, 
          {\tt PARSEC} \citep{Tang+2014, Chen+2015}, {\tt pyuvdata} \citep{pyuvdata, pyuvdata_Karto2025}, 
          {\tt SciPy} \citep{SciPy}, 
          {\tt Seaborn} \citep{Seaborn},
          {\tt ysoisochrone} \citep{Deng+2026}.
          }

\section*{Acknowledgments}
We thank the anonymous referee for their useful feedback and suggestions which improved the results and presentation of this manuscript.
J.B.L.~acknowledges the Smithsonian Institute for funding via a Submillimeter Array (SMA) Fellowship (2022--2025) and a CfA Prize Fellowship (J.C. Ryan; 2025--). 
R.A.~acknowledges funding from the Science \& Technology Facilities Council (STFC) through Consolidated Grant ST/W000857/1. 
M.C.W.~acknowledges funding from the Science \& Technology Facilities Council (STFC) through Grant UKRI1198. 
L.M. acknowledges funding by the European Union through the E-BEANS ERC project (grant No. 100117693). Views and opinions expressed are however those of the author(s) only and do not necessarily reflect those of the European Union or the European Research Council Executive Agency. Neither the European Union nor the granting authority can be held responsible for them.
The authors wish to recognize and acknowledge the very significant cultural role and reverence that the summit of Maunakea has always had within the indigenous Hawaiian community, where the Submillimeter Array is located. 
We are most fortunate to have the opportunity to conduct observations from this mountain.
We acknowledge the operational staff and scientists involved in the collection of data presented here.

This publication makes use of data products from the Wide-field Infrared Survey Explorer, which is a joint project of the University of California, Los Angeles, and the Jet Propulsion Laboratory/California Institute of Technology, funded by the National Aeronautics and Space Administration.
This publication makes use of data products from the Two Micron All Sky Survey, which is a joint project of the University of Massachusetts and the Infrared Processing and Analysis Center/California Institute of Technology, funded by the National Aeronautics and Space Administration and the National Science Foundation.
This work is based [in part] on observations made with the Spitzer Space Telescope, which is operated by the Jet Propulsion Laboratory, California Institute of Technology under a contract with NASA.

\section*{AI Disclosure Statement}
The AI tool Claude Sonnet 5 was used to review code used to generate a subset of figures in this manuscript, as well as review the population synthesis code. All outputs were verified by J.B.L.
No other AI tools were used in the generation of this manuscript or the analysis that this presents. 
Responsibility for the final manuscript lies entirely with the authors.
GAI tools are not listed as authors and do not bear responsibility for the final outcomes.
Declaration submitted by J.B.L.

\appendix{}
\section{Source properties and derived millimeter fluxes}
\label{sec:appData}
In this appendix we present the properties of the class~III YSOs with new SMA millimeter data.

\setlength{\LTcapwidth}{1.0\textwidth} 
\begin{longtable*}{l|cccccc}
   \caption{SMA observations table, showing source properties of those observed as part of this survey, ordered by Right Ascension (RA), and split into class~III YSOs (top half) and earlier-stage YSOs (bottom half). Notes: `R1' and `R2' refer to sources being members of Gaia-resolved multiples (with 1 and 2 designating the source being the primary or secondary member), `Sp' refers to sources that are spectroscopic binaries, `U' for unresolved binaries, and `Tri' for sources in stellar triples. Detections are presented with uncertainties that include an assumed flux calibration uncertainty of 10\% added in quadrature to the underlying fitting uncertainty (the significance of each detecion is presented in parantheses). Sources with names beginning `J' are shortened from `2MASS'. $^\dag$: The source HD~31305 was identified as a binary in \citet{Moor+2025} which we corroborate here, and report the properties of the secondary source HD~31305~B here. $^*$: SpTs estimated in this work. } \label{tab:mainTable} \\
    \hline
    \hline
    Name & RA & Decl. & SpT & YSO Class & Multiplicity & $F_{\rm 225\,GHz}$ \\
    & [deg] & [deg] & & & & [mJy] \\
    \hline
    \endhead

    \endfoot

    HD~285281 & 60.1295 & 19.5891 & F9 & III  & N/A & ${<}1.2$ \\
    BD+21~584 & 61.1187 & 21.9345 & F8 & III  & N/A & ${<}1.06$ \\
    HD~283323 & 61.3015 & 26.5454 & K2 & III  & N/A & ${<}1.0$ \\
    V1298~Tau & 61.3317 & 20.157 & K1.5 & III  & N/A & ${<}1.43$ \\
    HD~284149 & 61.6617 & 20.303 & G1 & III  & N/A & ${<}1.21$ \\
    V1195~Tau~B & 61.714 & 25.691 & K6.5 & III  & R2 & ${<}1.48$ \\
    V1195~Tau~A & 61.7141 & 25.6913 & K6.5 & III  & R1 & ${<}1.48$ \\
    HD~285458 & 61.9751 & 17.8403 & K4 & III  & N/A & ${<}1.47$ \\
    V1304~Tau & 62.3209 & 17.2689 & M1 & III  & N/A & ${<}3.96$ \\
    V1307~Tau & 63.2109 & 19.616 & K4.5 & III  & N/A & ${<}1.69$ \\
    HD~284266 & 63.8455 & 20.738 & K0 & III  & N/A & ${<}1.81$ \\
    V410~Tau & 64.6297 & 28.4544 & K3 & III  & U & ${<}1.25$ \\
    V1023~Tau & 64.696 & 28.3353 & K8.5 & III  & U & ${<}1.33$ \\
    V652~Per & 65.1005 & 31.3898 & K4.5 & III  & N/A & ${<}0.98$ \\
    HD~283572 & 65.4952 & 28.3017 & G4 & III  & N/A & ${<}0.97$ \\
    HD~27659~A & 65.7275 & 28.3983 & A3 & III  & R1 & ${<}1.46$ \\
    HD~27659~B$^*$ & 65.7278 & 28.3995 & G7 & III  & R2 & ${<}1.46$ \\
    HD~285751 & 65.9222 & 15.6319 & K2 & III  & N/A & ${<}1.42$ \\
    V1201~Tau & 66.2007 & 26.7211 & K1 & III  & N/A & ${<}1.71$ \\
    HD~283641 & 66.2044 & 26.7195 & K0 & III  & N/A & ${<}1.04$ \\
    BD+17~724~B & 66.7749 & 18.2102 & G5 & III  & N/A & ${<}1.15$ \\
    HD~285778 (HBC~388) & 66.794 & 17.8451 & G6 & III  & N/A & ${<}1.02$ \\
    HD~28354 & 67.3327 & 27.4041 & B9 & III  & N/A & ${<}0.77$ \\
    HD~284496 & 67.8202 & 21.8403 & K0 & III  & N/A & ${<}1.75$ \\
    L~1551-51 & 68.0387 & 17.9562 & K6 & III  & N/A & ${<}1.04$ \\
    HD~285840 & 68.1768 & 18.9194 & K1 & III  & N/A & ${<}1.62$ \\
    HD~28867~B & 68.3869 & 18.0167 & B9 & III  & N/A & ${<}0.67$ \\
    HD~28867~A & 68.3878 & 18.0166 & B9 & III  & N/A & ${<}1.26$ \\
    XEST~08-003 & 68.7373 & 22.9765 & M0 & III  & N/A & ${<}1.18$ \\
    HP~Tau/G3 & 68.973 & 22.9024 & M0.6 & III  & R, Tri & ${<}1.92$ \\
    HP~Tau/G2 & 68.9757 & 22.9037 & G2 & III  & R, Tri & ${<}1.92$ \\
    XEST~09-042 & 68.9956 & 22.6431 & K7 & III  & N/A & ${<}1.65$ \\
    V1115~Tau & 69.0796 & 25.7163 & K5 & III  & N/A & ${<}1.11$ \\
    V961~Per & 69.3203 & 31.1387 & K4 & III  & N/A & ${<}1.4$ \\
    RX~J0437.4+1851~A & 69.3621 & 18.8572 & K6 & III  & N/A & ${<}3.74$ \\
    V1325~Tau & 69.5544 & 20.3797 & K2 & III  & N/A & ${<}1.3$ \\
    V1117~Tau & 69.5651 & 23.0409 & M1 & III  & N/A & ${<}0.91$ \\
    V1326~Tau & 69.6152 & 15.7272 & K3 & III  & N/A & ${<}1.48$ \\
    HD~285957 & 69.6629 & 15.7703 & G6 & III  & N/A & ${<}1.48$ \\
    V583~Aur & 69.8561 & 33.5455 & K5 & III  & N/A & ${<}1.11$ \\
    HD~283798 & 70.4798 & 26.9803 & G7 & III  & N/A & ${<}1.1$ \\
    HD~283782 & 71.2269 & 27.2958 & K1 & III  & N/A & ${<}1.1$ \\
    COU~706~B$^*$ & 71.4603 & 28.6597 & G4 & III  & Sp & ${<}1.03$ \\
    HD~30171 & 71.4638 & 15.9304 & G3 & III  & N/A & ${<}1.36$ \\
    HD~30378 & 72.0948 & 29.7729 & B9.5 & III  & N/A & ${<}0.77$ \\
    V1341~Tau & 72.5008 & 22.4992 & K1 & III  & N/A & ${<}1.68$ \\
    V1346~Tau & 73.1281 & 17.5071 & K4 & III  & N/A & ${<}1.55$ \\
    V1347~Tau & 73.2091 & 16.3691 & K4.5 & III  & N/A & ${<}3.54$ \\
    V1348~Tau & 73.2378 & 19.3306 & K5 & III  & N/A & ${<}1.01$ \\
    HD~31281 & 73.7901 & 18.4419 & G1 & III  & N/A & ${<}3.87$ \\
    HD~282630 & 73.9041 & 30.2985 & K2 & III  & N/A & ${<}0.89$ \\
    HD~31305~A & 73.951 & 30.3378 & A1 & III  & U & ${<}0.87$ \\
    V1351~Tau & 74.0566 & 15.906 & K7 & III  & N/A & ${<}3.8$ \\
    HD~286179 & 74.2527 & 15.298 & G3 & III  & N/A & ${<}1.14$ \\
    HD~286178 & 74.3237 & 15.4193 & K1 & III  & N/A & ${<}1.46$ \\
    V1354~Tau & 74.3777 & 20.2415 & K3 & III  & N/A & ${<}1.84$ \\
    V1207~Tau & 74.6656 & 20.7788 & K7 & III  & N/A & ${<}1.78$ \\
    V1840~Ori & 74.9424 & 14.5153 & K4 & III  & N/A & ${<}3.67$ \\
    HD~31950 & 75.1013 & 15.0903 & F5 & III  & N/A & ${<}3.91$ \\
    LH98~157 & 75.1251 & 17.3997 & G5 & III  & N/A & ${<}1.75$ \\
    RX~J0507.2+2437 & 76.8003 & 24.6211 & K6 & III  & N/A & ${<}1.05$ \\
    2MASS~J05092844+2903003 & 77.3685 & 29.05 & K7 & III  & N/A & ${<}2.13$ \\
    LH98~175 & 77.794 & 28.2306 & K0 & III  & N/A & ${<}0.94$ \\
    LH98~185 & 80.7083 & 19.236 & M0 & III  & N/A & ${<}1.56$ \\
    V1364~Tau~A & 80.9777 & 25.5131 & K7 & III  & R1 & ${<}0.94$ \\
    V1364~Tau~B & 80.9785 & 25.5122 & M3 & III  & R2 & ${<}0.94$ \\
    LH98~190 & 81.1071 & 19.3685 & M0 & III  & N/A & ${<}1.62$ \\
    V1366~Tau & 82.427 & 23.5695 & K7 & III  & N/A & ${<}1.08$ \\
    LH98~211 & 84.1514 & 21.659 & K4 & III  & N/A & ${<}0.99$ \\
\hline
    IRAS~04278+2253~A & 67.7095 & 23.0024 & G8 & I  & R1 & 14.08 ${\pm}$ 1.7 (14.7\,$\sigma$) \\
    2MASS~J04333278+1800436 & 68.3868 & 18.012 & M1 & II  & N/A & 12.17 ${\pm}$ 1.34 (22.1\,$\sigma$) \\
    2MASS~J04355949+2238291 & 68.998 & 22.6413 & M1 & I  & N/A & 2.09 ${\pm}$ 0.63 (3.5\,$\sigma$) \\
    IRAS~04429+1550 & 71.464 & 15.9268 & M2.5 & II  & N/A & 54.7 ${\pm}$ 5.56 (54.7\,$\sigma$) \\
    HD~31305~B$^{\dag}$ & --- & --- &  K- & II  & U & 1.59 ${\pm}$ 0.37 (4.8\,$\sigma$) \\
\hline\end{longtable*}

\section{Source properties and derived thermal circumstellar dust excesses}
\label{sec:app_SED_XStable}
We present here the properties of the class~III YSOs that we found had significant thermal circumstellar dust excesses.

\begin{table*}[]
    \label{tab:SEDVals}
    \centering
    \caption{Table of YSOs determined to have with significant SED model fits to warm/cool ($T{<}500$\,K) component (16; top), a hot ($T{>}500$\,K) component (11; bottom). We present the best-fit values for the dust temperature $T_{\rm dust}$ and fractional luminosity $f$, and the implied blackbody radius $r_{\rm bb}$ from the dust temperature. The sources `2MJ0440', `2MJ0441', `2MJ05111', `2MJ0435', and 12MJ0421' are shortened for 2MASS~J04403979+2519061, 2MASS~J04414565+2301580, 2MASS~J05111053+2813504, 2MASS~J04354203+2252226, and 2MASS~J04210934+2750368 respectively. *: indicate sources with systematically bad parameter estimates.}

\begin{tabular}{l|cccc}
\hline
\hline
Name & SpT & $T_{\rm dust}$ & $f$ & $r_{\rm bb}$ \\
 & & [K] & $\times10^{-3}$ & [au] \\
\hline
HD~28867~A/B? & B9/B9 & 273${\pm}$20 & 0.331${\pm}$0.070 & 12.3${\pm}$2.2 \\
HD~30378 & B9.5 & 278${\pm}$63 & 0.058${\pm}$0.016 & 6.2${\pm}$2.9 \\
HD~27659~A & A3 & 218${\pm}$13 & 0.88${\pm}$0.14 & 6.6${\pm}$1.0 \\
2MJ05111 & K0 & 455${\pm}$61 & 0.28${\pm}$0.09 & 0.33${\pm}$0.04 \\
HD~286178 & K1 & 473${\pm}$133 & 1.77${\pm}$0.60 & 0.3${\pm}$0.1 \\
V1298~Tau* & K1.5 & 475${\pm}$320 & 0.22${\pm}$0.29 & 0.4${\pm}$0.8 \\
HD~282630 & K2 & 343${\pm}$23 & 1.33${\pm}$0.16 & 0.8${\pm}$0.1 \\
HBC~427 & K6 & 127${\pm}$13 & 1.17${\pm}$0.19 & 4.8${\pm}$1.0 \\
V1195~Tau~A/B? & K6.5/K6.5 & 250${\pm}$40 & 41.8${\pm}$8.9 & 1.2${\pm}$0.4 \\
V819~Tau & K8 & 144${\pm}$6 & 3.47${\pm}$0.13 & 2.5${\pm}$0.2 \\
V1320~Tau & K8 & 104${\pm}$5 & 4.02${\pm}$0.36 & 4.2${\pm}$0.4 \\
V827~Tau & M1.4 & 387${\pm}$92 & 0.37${\pm}$0.18 & 0.5${\pm}$0.2 \\
V1321~Tau & M1.5 & 161${\pm}$17 & 4.65${\pm}$0.53 & 1.8${\pm}$0.4 \\
2MJ0435* & M3 & 49${\pm}$19 & 5.29${\pm}$2.59 & 11.0${\pm}$10.0 \\
2MJ0441 & M4.3 & 172${\pm}$19 & 8.31${\pm}$1.12 & 0.9${\pm}$0.2 \\
2MJ0440 & M5 & 62${\pm}$10 & 91${\pm}$23 & 5.2${\pm}$1.9 \\
\hline
HD~285281 & F9 & 926${\pm}$46 & 1.69${\pm}$0.27 & 0.121${\pm}$0.012 \\
HD~31281* & G1 & 640${\pm}$310 & 0.22${\pm}$0.22 & 0.31${\pm}$0.46 \\
HD~285372 & K3 & 948${\pm}$39 & 2.84${\pm}$0.39 & 0.0569${\pm}$0.0048 \\
V1354~Tau & K3 & 930${\pm}$60 & 1.88${\pm}$0.47 & 0.083${\pm}$0.011 \\
V410~Tau & K3 & 978${\pm}$15 & 6.37${\pm}$0.63 & 0.128${\pm}$0.004 \\
L1551~51 & K6 & 941${\pm}$46 & 2.97${\pm}$0.55 & 0.062${\pm}$0.006 \\
V1076~Tau & K6 & 640${\pm}$60 & 2.96${\pm}$0.59 & 0.105${\pm}$0.019 \\
V826~Tau* & K7 & 620${\pm}$180 & 0.95${\pm}$0.66 & 0.18${\pm}$0.12 \\
FF~Tau* & K8 & 975${\pm}$17 & 14.0${\pm}$1.1 & 0.058${\pm}$0.002 \\
V999~Tau* & M0.6 & 890${\pm}$70 & 10.0${\pm}$2.0 & 0.066${\pm}$0.012 \\
2MJ0421 & M3.5 & 628${\pm}$37 & 47.8${\pm}$5.2 & 0.045${\pm}$0.006 \\
\hline
\end{tabular}
\end{table*}

\bibliography{sample631}{}
\bibliographystyle{aasjournal7.1}
\end{document}